\documentclass[times,doublespace]{simauth}

\usepackage{amsmath,amssymb,mathtools}
 {
      \theoremstyle{plain}
      
  }

\usepackage{xcolor}
\usepackage{graphicx,psfrag,epsf}
\usepackage{enumerate}
\usepackage[listings]{tcolorbox}
\usepackage{alltt}
\usepackage[french]{babel}

\usepackage{tikz}
\usetikzlibrary{arrows,shapes,trees}
\usepackage{multirow,floatrow,enumitem,subfloat,float}
\usepackage{lineno,blindtext,subcaption,setspace}
\usepackage{appendix,xr}
\usepackage{natbib,url}

    \usepackage{amsfonts}
    \usepackage{bm}

\usepackage{setspace}
\usepackage{amscd}
\usepackage{latexsym}
\usepackage[font={small,it}]{caption}
\usepackage{soul}

\usepackage[compact]{titlesec}
    \titlespacing{\section}{0pt}{2ex}{1ex}
    \titlespacing{\subsection}{0pt}{1ex}{0ex}
    \titlespacing{\subsubsection}{0pt}{0.5ex}{0ex}

\date{}

\newtheorem{theorem}{Theorem}

\def\b0{{0}}

\newcommand{\ts}{\texttt}

\graphicspath{{images/}} 

\usepackage[colorlinks,bookmarksopen,bookmarksnumbered,citecolor=red,urlcolor=red]{hyperref}

\newcommand\BibTeX{{\rmfamily B\kern-.05em \textsc{i\kern-.025em b}\kern-.08em
T\kern-.1667em\lower.7ex\hbox{E}\kern-.125emX}}

\def\volumeyear{2018}

\begin{document}

\runninghead{Xu \textit{et al}: MLMRT design for DIAMANTE study}

\title{Multi-Level Micro-Randomized Trial: Detecting the Proximal Effect of Messages on Physical Activity}

\author{
Jing Xu\affil{a}, 
Xiaoxi  Yan\affil{a}, 
Caroline Figueroa\affil{f},
Joseph Jay Williams\affil{c}, 
Bibhas Chakraborty\affil{a, d, e}
}

\address{
\affilnum{a} Centre for Quantitative Medicine, Duke-NUS Medical School, Singapore \\
\affilnum{c} Department of Computer Science, University of Toronto, Toronto, Canada\\
\affilnum{d} Department of Statistics and Applied Probability, National University of Singapore, Singapore\\
\affilnum{e} Department of Biostatistics and Bioinformatics, Duke University, Durham, NC, USA\\
\affilnum{f}  School of Social Welfare, University of California, Berkeley, USA\\
        }

\corraddr{ E-mail: \texttt{bibhas.chakraborty@duke-nus.edu.sg}}

\begin{abstract}
Technological advancements in mobile devices have made it possible to deliver mobile health interventions to individuals. 
A novel intervention framework that emerges from such advancements is the just-in-time adaptive intervention (JITAI), where it aims to suggest the right support to the individual ``just in time", when their needs arise, thus having proximal, near future effects.
The micro-randomized trial (MRT) design was proposed recently to test the proximal effects of these JITAIs. 
In an MRT, participants are repeatedly randomized to one of the intervention options of various in the intervention components, at a scale of hundreds or thousands of decision time points over the course of the study. 
However, the extant MRT framework only tests the proximal effect of two-level intervention components (e.g. control vs intervention).
In this paper, we propose a novel version of MRT design with multiple levels per intervention component, which we call ``multi-level micro-randomized trial” (MLMRT) design.  
The MLMRT extends the existing MRT design by allowing multi-level intervention components, and the addition of more levels to the components during the study period. 
We apply generalized estimating equation type methodology on the longitudinal data arising from an MLMRT to develop the novel test statistics for assessing the proximal effects and deriving the associated sample size calculators. 
We conduct simulation studies to evaluate the sample size calculators based on both power and precision. We have developed an R shiny application of the sample size calculators. This proposed design is motivated by our involvement in the Diabetes and Mental Health Adaptive Notification Tracking and Evaluation (DIAMANTE) study. 
This study uses a novel mobile application, also called ``DIAMANTE", which 
delivers adaptive text messages to encourage physical activity. 
\end{abstract}


\maketitle

\section{Introduction}\label{Intro}
Life is getting increasingly digital. 
The ubiquitous mobile devices are increasingly indispensable in our daily lives. 
They are not only used for communication, but are also used for a wide range of purposes such as calendar reminders, entertainment, activities tracking and health monitoring, among many others.
Mobile health or mHealth is a term used for the practice of medicine and health supported by mobile devices or wearables \cite{Sasan_2015}. They can better reach areas, people, and healthcare practitioners with limited exposure to certain aspects of healthcare. 
They are used across the health fields and include treatments of chronic disease (e.g., HIV \cite{Lewis_etal_2013}), increasing physical activity \cite{King_etal_2013}, supplement counseling or pharmacotherapy in treatment for substance use \cite{Marsch_2012}, and to support recovery from alcohol dependence \cite{Alessi_Petry_2013}.
Mobile interventions for anti-retroviral therapy and smoking cessation have shown sufficient effectiveness and replicability in trials and have been recommended for inclusion in health services \cite{Free_etal_2013}.

Mobile interventions do not only provide convenient health support, but can potentially be designed to deliver the support ``just in time" due to the increasingly ubiquitous access and powerful sensing capabilities of mobile devices and wearables. 
This novel intervention design is known as the ``just in time adaptive intervention" (JITAI), which aims 
to provide the right type or amount of support, at the right time (\cite{Intille_2004} and \cite{Patrick_etal_2008}), adapting according to an individual's changing internal and contextual state. 
The increasingly powerful mobile and sensing technologies underpins the use of JITAIs to support the health behaviors, due to an individual's state that can change rapidly and unexpectedly, in the natural environment. Nahum-Shani \textit{et al.} \cite{Nahum-Shani_etal_2018} bridge the gap between the growing technological capabilities for delivering JITAIs and research on the development and evaluation of these interventions.  

JITAI for mHealth is intended to have proximal or short term effects. A state-of-the-art experimental design proposed for testing the proximal effects of the intervention components in JITAIs is called the micro-randomized trial (MRT) design \cite{Klasnja_etal_2015}. The corresponding analysis plan and sample size calculation for current MRT designs are derived by Liao \textit{et al.} \cite{Liao_etal_2016}. 
In an MRT, there are numerous decision time points (up to hundreds or even thousands) for each participant throughout the study period. At each decision time point, the participant is randomly assigned to one of the available intervention options.
This sequential randomization captures the causal modelling of an intervention component’s time-varying proximal main effect as well as modelling of time-varying proximal moderation effect with another intervention component, similar to that of the factorial designs with multi-component interventions \cite{Chakraborty_etal_2009}. Note that the classical factorial designs do not consider the time-varying aspects in the main and moderating effects of the intervention components. 
There are several existing research studies using the MRT design, for example, `HeartSteps' for promoting physical activity (i.e. walking) among sedentary people \cite{Klasnja_etal_2019}; `Sense2Stop' for managing stress in newly abstinent smokers \cite{Liao_etal_2018}; and `JOOL' for engaging in self-monitoring using the proposed mHealth apps among office workers \cite{Bidargaddi_etal_2018}.  

Although the current MRT design allows for multiple intervention categries within an intervention component, its proximal effect is studied only at two levels (i.e. on versus off). This is akin to a two-level factorial design trial. 
For example, the HeartSteps study focuses on the proximal effect of active intervention message against no message. However, separately studying each of the intervention message levels may also be interesting. 
Rather than the proximal effect of the active intervention message levels, the proximal effect of an individual level within a component can be also recognized. 
This is similar to the multi-level factorial trials. 
A single multi-level trial can be more efficient and provide lower cost than using multiple two-level trials \cite{Ventz_etal_2017}.
Under a multi-level trial, it is also possible to propose new intervention levels during the study period as in plateform clinical trials \cite{Ventz_etal_2017}. 
The extension to a multi-level mHealth intervention message setting for MRTs is possible \cite{Boruvka_etal_2018}. 

In this paper, we propose a novel version of MRT design named \textit{multi-level micro-randomzsed trial} (MLMRT) design. 
Instead of estimating one proximal effect from each message component (two-level, i.e all the intervention message levels combined versus the control level), the proximal effects of individual message levels within each component versus no message (control) can be estimated. 
In practice, there is a possibility that the effective message levels are not all proposed at the beginning of the study \cite{Lee_etal_2019}. 
In this paper, we also extend the flexibility of the design by allowing new proposed intervention message levels to be added later during the study period. 

Novel test statistics are derived to detect the proximal effects from MLMRT data. 
We present the corresponding sample size calculators so that the trials can be sized to either detect the standardized proximal effect sizes at a nominal power, or estimate the standardized proximal effect sizes within a margin of error (e.g., the half-width of the confidence interval), for example, when the prior infromation of the proximal effect is unknown, therefore a pilot study can be recommended.
On top of the constant, linear and quadratic trends for the proximal effect proposed by Liao et al (2016) \cite{Liao_etal_2016} for the HeartSteps study, the proposed MLMRT design also considers the mixture of linear and constant trends (similar to a linear spline). This combination of linear and constant trends captures that the proximal effect is improved until reaching, then maintaining the maximum value.

This work is motivated by our collaboration on the Diabetes and Mental Health Adaptive Notification Tracking and Evaluation (DIAMANTE) study \textit{et al.} \cite{Aguilera_etal_2020}. We employ the proposed MLMRT design, the corresponding analysis plan and associated sample size calculators to this study. The DIAMANTE study uses an mHealth app named ``DIAMANTE''. 
This app implements a text-messaging platform `HealthySMS' used to send text messages in smartphones in order to encourage physical activity (i.e. walking) among comorbid diabetes and depression patients, from low-income and ethnic minority families served in the San Francisco Health Network. 
Unlike the HeartSteps study, our study includes not only an uniform random intervention (URI) group, but also an adaptive intervention (AI) group.
The participants of URI group receive the timing and message types with equal probabilty while the participants of AI group receive
the timing and message types learnt adaptively by a reinforcement learning algorithm in order to reach and maintain the maximum value of the proximal effects.
The purpose of this study is to understand how text message programs work and also to learn the best ways to develop and deliver text messages to motivate individuals to do regular exercise for maintaining both physical and mental well-being in daily life.
However, the proposed MLMRT design can be applied to both the URI and AI groups. Note that the proposed sample size calculator is only applied to the URI group.

Therefore we summarize four novelties of the proposed method. 
First, we extend the MRT designs that consider the intervention components with multiple levels.
Second, some levels are allowed to be added later during the study period.
Third, the sample size calculators of the proposed design are derived by not only the power-based method, but also the precision-based method.
Last, the proposed method is applied to a new study named ``DIAMANTE''. This study develops a novel app also named ``DIAMANTE''. This app  delivers the text messages through the smartphones. The messages can be learnt by the reinforcement learning algorithm in order to maintain the best physical activity outcomes.  

The rest of the paper is organised as follows. 
Section \ref{s:analysissamplesize} describes the proposed MLMRT design, the statistical models and the sample size calculation methods for our motivating study called ``DIAMANTE''.
Section \ref{s:sim} investigates the finite-sample performance of the proposed sample size calculation methods through simulation studies. Section \ref{s: rimp} demonstrates the implementation of the corresponding R function `SampleSize\_MLMRT' (available at the GitHub link {\url{https://github.com/Kenny-Jing-Xu/MLMRT-SS/blob/master/SampleSizeMLMRT.R}) and R shiny app `MLMRT-SS' (\url{https://kennyxu.shinyapps.io/mlmrt_shinyapps/}). 
Section \ref{s:realexample} demonstrates the proposed MLMRT sample size calculator for DIAMANTE study through a data analysis of a pilot study.
Finally, the paper ends with a discussion in Section \ref{s:disc}. 
Supplementary materials, consisting of detailed derivations and additional simulation results are relegated to the Appendix.

\section{MLMRT Design for DIAMANTE: Statistical Model Estimation and Sample Size Calculation}\label{s:analysissamplesize}
In this section, we describe the proposed MLMRT design, and the corresponding analysis plan and sample size calculation methods for the DIAMANTE study.

\subsection{Multi-Level Micro-Randomized Trial Design}\label{s:MLMRT}
Our motivating study is named ``DIAMANTE'' (Diabetes and Mental Health Adaptive Notification Tracking and Evaluation). 
It develops a mobile health application, which is also named `DIAMANTE', see Figure \ref{fig: DIAMANTE}. The web link of this mobile application is \url{https://diamante.healthysms.org}. It uses mobile device technology to deliver physical activity interventions, which focus on the cohort of patients with comorbid diabetes and depression from low-income and ethnic minority families in the San Francisco Health Network. 
The study is a three-arm randomised clinical trial, where participants are randomized to receive either the control, uniform random, or adaptive intervention of message program.
Though the proposed multi-level micro-randomised trial (MLMRT) design can be applied to either the uniform random intervention(URI) or adaptive intervention (AI) group, the proposed sample size calculator is applied to the URI group.

The MLMRT design employed in the DIAMANTE study has three multi-level components. They are `Time Window', `Feedback Message' and `Motivational Message'. Each participant receives one of the intervention messages in `Feedback Message' and `Motivational Message' at most one time per day at one of the timing levels in the `Time Window'. 
The two different messages are one minute apart. 
The primary outcome measure is the walking steps counted in the next 24 hours after the messages (either the control or intervention levels) sent. The study period will be six month long with one decision time point per day, so the participants could be randomised approximately 180 times in the study. 

The Time Window component (four-level, i.e. $A_{\textbf{T}}$=0, 1, 2, 3) suggests when to send the messages, i.e. level-0 (9:00-11:30am), level-1 (11:30am-14:00pm), level-2 (14:00-16:30pm) and level-3 (16:30-19:00pm). A participant is randomly assigned to one of the time window on each day of the study period.

The Feedback Message  (five-level, i.e. $A_{\textbf{F}}$=0, 1, 2, 3, 4) has reference level-0 (i.e. no message or control). It has four active intervention levels,  i.e. level-1 (reaching goal, e.g., a message like ``Yesterday, you did not reach
your goal."), level-2 (steps walked yesterday, e.g., a message like ``Yesterday, you walked 3824 steps."), level-3 (walked more or less yesterday than day before, e.g., a message like ``Yesterday, you walked more than day before."), level-4 (steps walked yesterday plus a motivational message, e.g., a message like ``You walked 8000 steps yesterday. Great job!").

Similar to the  Feedback Message, the Motivational Message (four-level, i.e. $A_{\textbf{M}}$=0, 1, 2, 3) has reference level-0 (i.e. no message or control). It has three active intervention levels, i.e. level-1 (benefit, e.g., a message like ``Doing more physical activity can help reduce feelings of fatigue."), level-2 (self-efficacy, e.g., a message like ``You have made changes to improve your health before, you can do it again.") and level-3 (opportunity, e.g., a message like ``It there a local park you have been waiting to visit?
Use it as an opportunity to get out of the house and do more steps!"). 

The message components within the adaptive intervention use the adaptive mobile messaging technologies, i.e. daily messages learned by reinforcement algorithm \cite{Yom-Tov_etal_2017}, to encourage healthy living and positive behaviour change in physical and mental health settings. The details are covered by the protocol paper \cite{Aguilera_etal_2020}. The purpose of the adaptive intervention group is to learn the best ways to develop and deliver motivational and feedback text messages according to the current context of the participants to encourage regular physical activity in daily life. 

The MLMRT is a cutting-edge trial design suitable to take care of the time-varying, sequential nature of the multi-component interventions with multiple levels, akin to a sequential multi-level full-factorial design. At each day of the study period, each participant is randomized to one of the levels for each component of Time Window, Feedback Message and Motivational Message. 
By considering the proximal effect of Feedback Message component, the primary outcome is the difference in steps counted between the control level and either of the interventional levels in the next 24-hour after the messages are sent ($Y$). The generalised estimating equation method \cite{Liang_Zeger_1986} is applied for longitudinal data analysis and sample size calculation, where some additional intervention message levels can be considered to be included half-way through the study period.
This adds more flexibility to the design in case new promising messages become available after the onset of the study, and is thus similar in spirit to platform clinical trials.

We consider the long format of the longitudinal dataset from this study. Each row represents a participant at a particular day of the study, with the dummy variables $A_{\text{T1}}$, $\ldots$, $A_{\text{T3}}$, $A_{\text{F1}}$, $\ldots$, $A_{\text{F4}}$, $A_{\text{M1}}$, $\ldots$, $A_{\text{M3}}$, and the proximal response $Y$. $A_{\text{F1}}$, $\ldots$, $A_{\text{F4}}$ are the dummy variables of $A_{\textbf{F}}$ and so on, such that if the participant on that day receives control (or level-0) from the Feedback Message component, then the indicators give $A_{\text{F1}}=\cdots=A_{\text{F4}}$=0 .


There is a pilot study from the DIAMANTE project. It includes 84 undergraduate and postgraduate students from the University of California, Berkeley. 
Only 22 of them received adaptive intervention of message program while the rest received uniform random intervention of message program. The study period is 45 days. 
The primary outcome measure of this pilot study is the change in walking steps counted, i.e. the day of intervention messages sent minus the day before. 
In Section \ref{s:realexample}, we use the data analysis results of this student dataset to demonstrate the proposed sample size calculator for the MLMRT design.  

\begin{figure}[H]
\begin{center}
\begin{tabular}{c}
 \includegraphics[scale=0.3]{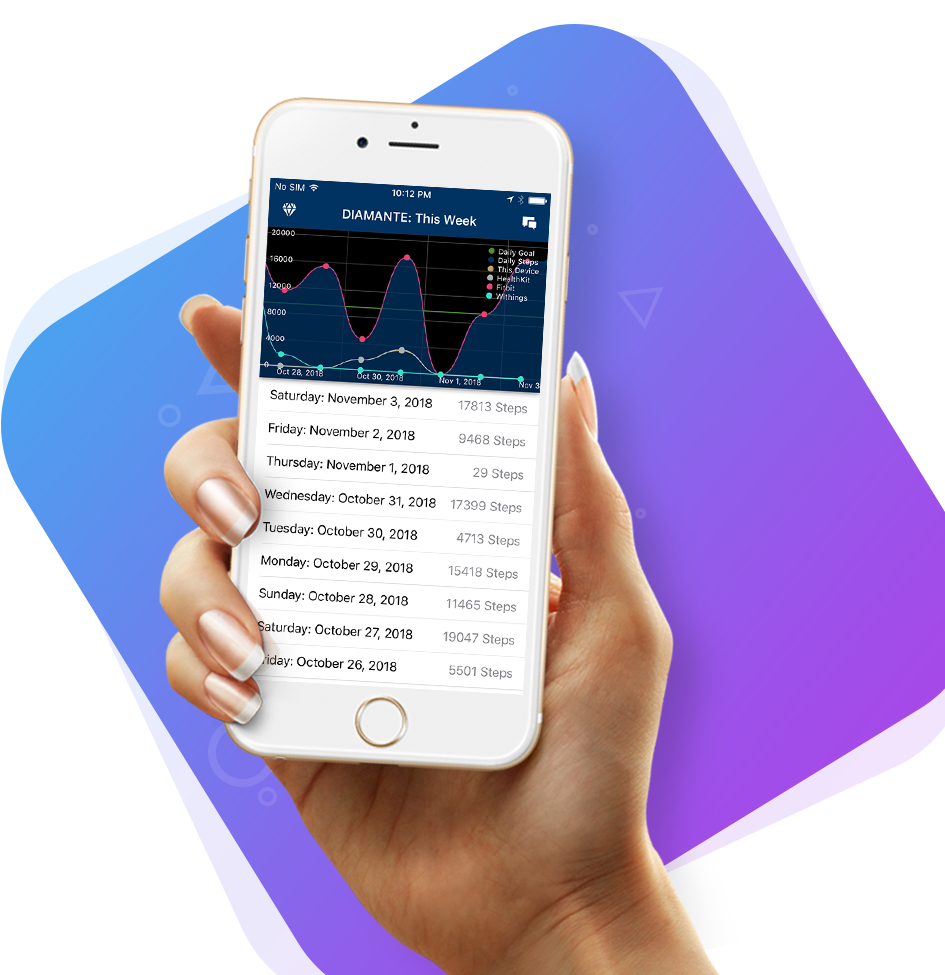} 
\end{tabular}
\end{center}
\caption{
Diabetes and Mental Health Adaptive Notification Tracking and Evaluation
}\label{fig: DIAMANTE}
\end{figure}

\subsection{Statistical Model}\label{s:analysis}
In this section, we extend the statistical model proposed by \cite{Liao_etal_2016} for a micro-randomization trial (MRT) \cite{Klasnja_etal_2015} to multi-level components. 
Instead of estimating the proximal effect of the intervention levels combined, we estimate the individual effects for each level. 
We do not restrict the assumption of fixed allocation probability of each level at each decision time point. 
Furthermore, similar way to adding arm platform clinical trials \cite{Ventz_etal_2017}, we allow some of the new intervention levels to be added during the study period based on the idea of crowdsourcing, e.g., De Vries et al (2016) \cite{DeVries_etal_2016}.
The crowdsourcing can be an effective method to propose new message levels to the participants according to their stages changed.

In order to test the proximal effect of a particular component, our null hypothesis is 
\begin{equation*}
H_{0}: \beta_1(d)=\cdots =\beta_{M_0}(d)=\cdots =\beta_{M_0+M_1}(d)=\cdots=\beta_{\sum_{j=0}^{k}M_j}(d)=0, 
\end{equation*}
with $d=1,\ldots,D$ (the study period in days); $M_0$ is the number of message levels initially at day $d_0$ (e.g., $d=1$); $M_1$ is  number of message levels added at day $d_1$ (first adding time),$\ldots$, $M_k$ is the number of message levels added at the last adding day $d_k$, where $d_0\leq 1<d_1\ldots <d_k\leq D$. We propose an alternative hypothesis similar to Liao et al (2016) \cite{Liao_etal_2016}
\begin{equation*}
H_{1}: \beta_1(d)=\boldsymbol Z_{d}^{\top}\boldsymbol\beta_1~\text{or}~ \cdots \beta_{M_0}(d)=\boldsymbol Z_{d}^{\top}\boldsymbol\beta_{M_0}~\text{or}~\cdots \beta_{M_0+M_1}(d)=\boldsymbol Z_{d}^{\top}\boldsymbol\beta_{M_0+M1}~\text{or}~\cdots\beta_{\sum_{j=0}^{k}M_j}(d)=\boldsymbol Z_{d}^{\top}\boldsymbol\beta_{\sum_{j=0}^{k}M_j}. 
\end{equation*}
Suppose the alternative is of $(p-1)^{\text{th}}$ order, i.e. 
\begin{equation}\label{Zbeta_p}
Z_{d}^{\top}\boldsymbol\beta_m = \left( 1, \ldots, ( d - 1 )^{p-1} \right)\boldsymbol\beta_m,
\end{equation}
where $m$ indicates the $m^{\text{th}}$ intervention message level, i.e. $m=1,\ldots,M_0,\ldots,M_0+M_1,\ldots,\sum_{j=0}^{k}M_j$, $\boldsymbol\beta_m=(\beta_{m1},\ldots, \beta_{mp})^{\top}$. The control level is represented by $m=0$. 
Similar to the MRT design of Liao et al (2016) \cite{Liao_etal_2016}, we have $p=$1, 2 and 3 for the constant, linear and quadratic trends of the proximal effect respectively. 
However, for the combination of linear and constant trend, which captures the proximal effect is improved until reaching the maximum value at day $d_{\text{max}}$, then maintain the maximum value afterwards.
Therefore, the alternative can be also defined by a linear spline, i.e.
\begin{equation}\label{Zbeta_lspline}
Z_{d}^{\top}\boldsymbol\beta_m =  \left( 1, \text{min}\left[d_{\text{max}} -1, d - 1 \right] \right).
\end{equation}

Note that $\boldsymbol Z$ of equation (\ref{Zbeta_p}) is restricted to one decision time point per day (that is consistent to the DIAMANTE study) in this paper. However it can be extended to a more general form by involving the number of decision time points $T$ per day, i.e.
\begin{equation*}
Z_{dt}^{\top}\boldsymbol\beta_m = \left( 1, \left[ \dfrac{( d - 1 )T+t-1}{T} \right], \ldots, \left[  \dfrac{( d - 1 )T+t-1}{T} \right]^{p-1} \right)\boldsymbol\beta_m,
\end{equation*}
where $t=1, \ldots, T$. This form is considered in the R function `SampleSize\_MLMRT', see Sections \ref{s:demo} and \ref{s: Rshiny}.

A simple regression model considering the proximal effect of a component based on the alternative hypothesis can be written. First, we let the proximal response be $Y_{id}$, and all the intervention message levels be $\boldsymbol A_{id}$=($A_{i1d}$,$\ldots$,$A_{iM_0d}$,$\ldots$,$A_{i( \sum_{j=0}^{k-1}M_j+1 )d}$,$\ldots$,$A_{i( \sum_{j=0}^{k}M_j )d}$)$^{\top}$ for participant $i$ at day $d$. The probabilities of participant $i$ receiving $\boldsymbol A_{id}$ at day $d$ follows a multinomial distribution, Multinomial(1-$\sum_m\pi_{md}$, $\pi_{1d}$,$\ldots$,$\pi_{( \sum_{j=0}^{k}M_j  )d}$), where $\pi_{(\cdot)}$ is the randomization probability corresponding to $A_{i(\cdot)}$.
It should be note that when $d < d_1$, new intervention message levels $A_{ i ( M_0 + 1 ) d }$,$\ldots$,$A_{ i ( M_0 + M_1 ) d }$, are not yet available, so the corresponding randomization probabilities are zeros. We then denote the proximal effects of all intervention message levels at day $d$ to be $\boldsymbol Z_{d}^{\top}\boldsymbol\beta$=($\boldsymbol Z_{d}^{\top}\boldsymbol\beta_1$, $\ldots$, $\boldsymbol Z_{d}^{\top}\boldsymbol\beta_{M_0}$,$\ldots$, $\boldsymbol Z_{d}^{\top}\boldsymbol\beta_{\sum_{j=0}^{k-1}M_j+1}$, $\ldots$, $\boldsymbol Z_{d}^{\top}\boldsymbol\beta_{\sum_{j=0}^{k}M_j}$)$^{\top}$.
 The working model is then written as 
\begin{align*}
Y_{id}=&\boldsymbol B_{d}^{\top}\boldsymbol\alpha\\
&+\left[ (A_{i1d}-\pi_{1d})\boldsymbol Z_{d}^{\top}\boldsymbol\beta_1+ \cdots +(A_{iM_{0}d}-\pi_{M_{0}d})\boldsymbol Z_{d}^{\top}\boldsymbol\beta_{M_0} \right]\\
&+\left[ (A_{i( M_0 + 1 )d}-\pi_{( M_0 + 1 )d})\boldsymbol Z_{d}^{\top}\boldsymbol\beta_{ ( M_0 + 1 ) }+ \cdots +(A_{i( M_{0} + M_{1} ) d}-\pi_{( M_{0} + M_{1} )d})\boldsymbol Z_{d}^{\top}\boldsymbol\beta_{( M_{0} + M_{1} )} \right]\\
&\vdots\\
&+\left[ (A_{i( \sum_{j=0}^{k-1} M_j +1) d}-\pi_{( \sum_{j=0}^{k-1}M_j + 1 )d})\boldsymbol Z_{d}^{\top}\boldsymbol\beta_{ ( \sum_{j=0}^{k-1}M_j + 1 ) }+ \cdots +(A_{i( \sum_{j=0}^{k}M_j ) d}-\pi_{( \sum_{j=0}^{k}M_j )d})\boldsymbol Z_{d}^{\top}\boldsymbol\beta_{( \sum_{j=0}^{k}M_j )} \right]\\
&+\epsilon_{id}\\,
\end{align*}
where $\boldsymbol B_{d}^{\top}\boldsymbol\alpha$ is the intercept term with $\boldsymbol B_{d}^{\top}$ = $\left( 1, d - 1, \ldots, ( d - 1 )^{q-1} \right)$, for a $(q-1)^{\text{th}}$-order and $\boldsymbol\alpha^{\top}$=$(\alpha_1, \ldots, \alpha_q)$; $\boldsymbol\beta_{m}^{\top}$=($\beta_{m1},\ldots,\beta_{mp})$ for the $m^{\text{th}}$ intervention message level; and  $\epsilon_{id}$ are independently and identically distributed (i.i.d) with zero mean and constant variance.

We define $\boldsymbol\theta$=($\boldsymbol\alpha^{\top}$, $\boldsymbol\beta^{\top}$)$^{\top}$. The estimator $\hat{\boldsymbol\theta}$ is obtained by minimizing the least squares error,
\begin{equation}
LSE(\boldsymbol\theta)=\dfrac{1}{N}\sum_{i=1}^{N}\sum_{d=1}^{D}I_{id}\left[ Y_{i, d}- \boldsymbol X_{id}^{\top}\boldsymbol\theta \right]^2,
\end{equation} 
where `N' is the sample size, `D' is the study period in days; $I_{id}$ is the availability indicator for participant $i$ at day $d$, i.e. $E(I_{id})=\tau_{d}$. 
Note that our DIAMANTE study assumes the participants are always available (i.e. $\tau_{d}=1$), whereas other studies (e.g., HeartSteps) does not. Hence, in the similar way to the MRT platform \cite{Liao_etal_2016}, we also consider the availabilities for the parameters estimation. 
We have 
\begin{equation*}
\dfrac{\partial}{\partial\boldsymbol\theta}LSE(\boldsymbol\theta)=\dfrac{-2}{N}\sum_{i=1}^{N}\sum_{d=1}^{D}I_{id}\left[ Y_{i, d}-\boldsymbol X_{id}^{\top} \boldsymbol\theta \right]\boldsymbol X_{id}=0,
\end{equation*}
and
\begin{equation}\label{theta_hat}
\hat{\boldsymbol\theta}=\left( \dfrac{1}{N}\sum_{i=1}^{N}\sum_{d=1}^{D}I_{id}\boldsymbol X_{id} \boldsymbol X_{id}^{\top} \right)^{-1}\dfrac{1}{N}\sum_{i=1}^{N}\sum_{d=1}^{D}I_{id}Y_{i, d}\boldsymbol X_{id}.
\end{equation}
Let $\hat{\boldsymbol\theta}\rightarrow\tilde{\boldsymbol\theta}$ when $N\rightarrow\infty$, such that 
\begin{equation}\label{theta_tilde_aa}
\tilde{\boldsymbol\theta}=\left[\sum_{d=1}^{D}E\left( I_{id} \boldsymbol X_{id} \boldsymbol X_{id}^{\top} \right)\right]^{-1} \sum_{d=1}^{D}E\left( I_{id}Y_{i, d}\boldsymbol X_{id} \right).
\end{equation}
More details about equation (\ref{theta_tilde_aa}) are covered in Appendix. 

In this section, we simplify the form of the working model by making the following assumptions. 

\begin{enumerate}
\item[a)] $E(Y_{i, d}\mid I_{id}=1)$=$\boldsymbol B_{d}^{\top}\boldsymbol\alpha$, for some $\boldsymbol\alpha\in R^{q}$.
\item[b)] $\beta_m(d)$=$\boldsymbol Z_{d}^{\top}\boldsymbol\beta_{m}$ for $m=1,\ldots,M_0, \ldots,\sum_{j=0}^{k}M_j$, $\boldsymbol\beta_{m}\in R^{p}$.
\item[c)] $\text{Var}(Y_{i, d}\mid I_{id}=1, A_{imd})$ is constant in $d$ for $m=1,\ldots,M_0,\ldots,\sum_{j=0}^{k}M_j$.
\item[d)] $E[\tilde{\epsilon}_{i(d)}\tilde{\epsilon}_{i(d)^\prime}\mid I_{i(d)}=1, I_{i(d)^\prime}=1, A_{im(d)}, A_{im(d)^\prime}]$ is constant in in $d$ for $m=1,\ldots,M_0,\ldots,\sum_{j=0}^{k}M_j$.
\item[e)] $\boldsymbol Y_{i}$=$(Y_{i1},\ldots, Y_{iD})^{\top}$ are independent and identically distributed; $I_{id}$, $\tilde{\epsilon}_{id}$ and $\boldsymbol X_{id}$ are independent and their distributions are independent of $\boldsymbol\theta$.
\item[f)] Let $\boldsymbol\Theta$ be the parameter space for $\boldsymbol\theta$, where $\boldsymbol\Theta$ is a compact subset of $R^{q+p\sum_{j=0}^{k}M_j}$. 
\item[g)] $E(LSE(\boldsymbol\theta))$ exists and has a zero value at $\tilde{\boldsymbol\theta}\in\boldsymbol\Theta$. 
\item[h)] $LSE(\boldsymbol\theta)$  is continuous, bounded and differentiable at neighbourhood of $\tilde{\boldsymbol\theta}$.
\item[i)] The matrix $\sum_{d=1}^{D}E\left( I_{id} \boldsymbol X_{id} \boldsymbol X_{id}^{\top} \right)$ in equation (\ref{theta_tilde_aa}) is invertible.
\end{enumerate}

\subsection{Test Statistics}\label{test}
This section presents and derives the test statistic distributions. First we present a following theorem of consistency and asymptotic normality for the least squares estimator $\hat{\boldsymbol\theta}$. The proof is covered in Appendix.
\begin{theorem}\label{theo1}
The least squares estimator $\hat{\boldsymbol\theta}$ is a consistent estimator of $\tilde{\boldsymbol\theta}$. Under moment conditions and the Assumptions a)-i), we have $\sqrt{N}(\hat{\boldsymbol\theta}-\tilde{\boldsymbol\theta})\rightarrow \text{Normal}(0, \boldsymbol\Sigma_{\boldsymbol\theta})$.
\end{theorem}
The asymptotic covariance matrix  
$\boldsymbol\Sigma_{\boldsymbol\theta}$ is defined by
\begin{align*}
\left[ \sum_{d=1}^{D}E( I_{id}\boldsymbol X_{id} \boldsymbol X_{id}^{\top}) \right]^{-1} 
E\left(\sum_{d=1}^{D} I_{id}\tilde{\epsilon}_{id}\boldsymbol X_{id}  \sum_{d=1}^{D} I_{id}\tilde{\epsilon}_{id}\boldsymbol X_{id}^{\top} \right) \left[ \sum_{d=1}^{D}E( I_{id}\boldsymbol X_{id} \boldsymbol X_{id}^{\top}) \right]^{-1},
\end{align*} 
where $E\left(\sum_{d=1}^{D} I_{id}\tilde{\epsilon}_{id}\boldsymbol X_{id}  \sum_{d=1}^{D} I_{id}\tilde{\epsilon}_{id}\boldsymbol X_{id}^{\top}  \right)$ is defined in the Appendix.
Thus the asymptotic distribution of $\hat{\boldsymbol\beta}$ converges to normal, i.e.
$\sqrt{N}(\hat{\boldsymbol\beta}-\tilde{\boldsymbol\beta})\rightarrow Normal(0, \boldsymbol\Sigma_{\boldsymbol\beta\boldsymbol\beta})$ with covariance matrix $\boldsymbol\Sigma_{\boldsymbol\beta\boldsymbol\beta}$; see Appendix for its derivation and we have $\bar{\sigma}^2$=$\sum_{d=1}^{D}E(\text{Var}(Y_{i, d}\mid I_{id}=1, A_{i1d},\ldots,A_{i \sum_{j=0}^{k}M_j d} ))/(D)$. 
The asymptotic covariance matrix can be expressed by $\boldsymbol\Sigma_{\boldsymbol\beta\boldsymbol\beta}$=$\boldsymbol Q^{-1} \boldsymbol W \boldsymbol Q^{-1}$. 

Next we define test statistics $\hat{C}_N(\boldsymbol\delta)$=$N\hat{\boldsymbol\beta}^{\top}\boldsymbol\Sigma_{\boldsymbol\beta\boldsymbol\beta}^{-1}\hat{\boldsymbol\beta}$, which follows a
$\chi^2( ( \sum_{j=0}^{k}M_j ) p)$ distribution when the null hypothesis is true and $N$ is large \cite{Tu_etal_2004}, where $\boldsymbol\delta=\boldsymbol\beta/\sigma$. 
Tu \textit{et al.} \cite{Tu_etal_2004} also proposes sample size calculation method based on power under GEE approach for longitudinal data. Hence we can define the power function as follow. If $H_0$: $\boldsymbol\beta=0$ is true, and the type-I error rate is defined by
\begin{equation}\label{N_chi_alpha}
\text{Pr}\left(X_{ ( \sum_{j=0}^{k}M_j )p )} > \chi_{( \sum_{j=0}^{k}M_j )p, \alpha}\right)=\alpha,
\end{equation}
where $X_{ ( \sum_{j=0}^{k}M_j )p )}$ presents a random variable that follows a centrality Chi-squared distribution with degrees of freedom $p\sum_{j=0}^{k}M_j$ while $\chi_{( \sum_{j=0}^{k}M_j )p, \alpha}$ represents its $1-\alpha$ quantile. 
We reject $H_0$ at $5\%$ level of significance if $X_{ ( \sum_{j=0}^{k}M_j )p )} > \chi_{( \sum_{j=0}^{k}M_j )p, \alpha}$.
If $H_1$: $\boldsymbol\beta=\tilde{\boldsymbol\beta}\neq 0$ is true, then
\begin{equation}\label{N_chi_P}
\text{Pr}\left(X_{ ( \sum_{j=0}^{k}M_j )p, \tilde{C}_N(\boldsymbol\delta )} > \chi_{( \sum_{j=0}^{k}M_j )p, \alpha}\right)=\text{Power},
\end{equation}
where $X_{ ( \sum_{j=0}^{k}M_j )p,  \tilde{C}_N(\boldsymbol\delta)}$ represents a random variable that follows a Chi-squared distribution with degrees of freedom $p\sum_{j=0}^{k}M_j$ and non-centrality parameter  $\tilde{C}_N(\boldsymbol\delta)$. Note that \cite{Tu_etal_2004} does not define the distribution of $\hat{C}_N(\boldsymbol\delta)$ for a small sample size.

When $N$ is small, $\boldsymbol\Sigma_{\boldsymbol\beta\boldsymbol\beta}$ is replaced by $\hat{\boldsymbol\Sigma}_{\boldsymbol\beta\boldsymbol\beta}$, which is derived by \cite{Mancl_DeRouen_2001} 
, and the test statistic follows Hotelling's T-squared distribution (e.g., see \cite{Hotelling_1931} and \cite{Li_Redden_2015}). 
We define the small sample estimator by $\hat{\boldsymbol\Sigma}_{\boldsymbol\beta\boldsymbol\beta}$=$\hat{\boldsymbol Q}^{-1} \hat{\boldsymbol W} \hat{\boldsymbol Q}^{-1}$. Let
\begin{equation}
\hat{e}_{id}=Y_{i, d} - \boldsymbol X_{id}^{\top}\hat{\boldsymbol\theta}
\end{equation}
\begin{equation}
\hat{\boldsymbol e}_{i}^{\top}=(\hat{e}_{i1}, \ldots, \hat{e}_{iD}),
\end{equation}
\begin{equation}
\boldsymbol X_{i}^{\top}=
\begin{bmatrix}
\boldsymbol X_{i1}^{\top}I_{i1}\\
\vdots \\
\boldsymbol X_{iD}^{\top}I_{iD}
\end{bmatrix}_{(D)\times (q+ ( \sum_{j=0}^{k}M_j )p)}
\end{equation}
\begin{equation}
\boldsymbol H_{i}=\boldsymbol X_{i}^{\top}[\sum_{i=1}^N \boldsymbol X_{i} \boldsymbol X_{i}^{\top}]^{-1} \boldsymbol X_{i}.
\end{equation}
The matrix $\hat{\boldsymbol Q}^{-1}$ is given by the lower right $( \sum_{j=0}^{k}M_j )p\times ( \sum_{j=0}^{k}M_j )p$ block of $[\sum_{i=1}^N \boldsymbol X_{i} \boldsymbol X_{i}^{\top}/N]^{-1}$; the matrix $\hat{\boldsymbol W}$ is given by the lower right $( \sum_{j=0}^{k}M_j )p\times ( \sum_{j=0}^{k}M_j )p$ block of $[\sum_{i=1}^N \boldsymbol X_{i} (\boldsymbol I_{D\times D} - \boldsymbol H_i)^{-1} \hat{\boldsymbol e}_{i}\hat{\boldsymbol e}_{i}^{\top} (\boldsymbol I_{D\times D} - \boldsymbol H_i)^{-1} \boldsymbol X_{i}^{\top}]/N$, where $\boldsymbol I_{D\times D}$ is the identity matrix with dimension of $D\times D$. 

Liao \textit{et al.} \cite{Liao_etal_2016} suggests that the test statistic follows a Hotelling's $T^2$ distribution with dimension of $p\sum_{j=0}^{k}M_j$ and degrees of freedom $N-q-1$ approximately, i.e.
\begin{align*}
\hat{C}_N(\boldsymbol\delta)=N\hat{\boldsymbol\beta}^{\top}\hat{\boldsymbol\Sigma}_{\boldsymbol\beta\boldsymbol\beta}^{-1}\hat{\boldsymbol\beta}\sim T^2_{ ( \sum_{j=0}^{k}M_j ) p, N-q-1}=\dfrac{ ( \sum_{j=0}^{k}M_j ) p(N-q-1)}{N-q- ( \sum_{j=0}^{k}M_j ) p} F_{ ( \sum_{j=0}^{k}M_j ) p, N-q- ( \sum_{j=0}^{k}M_j ) p},
\end{align*} 
Thus,
\begin{align*}
\dfrac{N-q- ( \sum_{j=0}^{k}M_j ) p}{ ( \sum_{j=0}^{k}M_j ) p(N-q-1)} \hat{C}_N(\boldsymbol\delta)\sim F_{ ( \sum_{j=0}^{k}M_j ) p, N-q- ( \sum_{j=0}^{k}M_j )p},
\end{align*}
if $H_0$: $\boldsymbol\beta=0$ is true, and the type-I error ($\alpha$) is defined by
\begin{equation}\label{N_F_N_q_1_alpha}
\text{Pr}\left(F_{ ( \sum_{j=0}^{k}M_j )p, N-q-( \sum_{j=0}^{k}M_j )p} > f_{( \sum_{j=0}^{k}M_j )p, N-q-( \sum_{j=0}^{k}M_j ) p, \alpha}\right)=\alpha.
\end{equation}
where $F_{ ( \sum_{j=0}^{k}M_j )p,  N-q-( \sum_{j=0}^{k}M_j )p )}$ represents a random variable that follows a centrality $F$ distribution with degrees of freedom $p\sum_{j=0}^{k}M_j$ and $N-q-( \sum_{j=0}^{k}M_j )p$ while $f_{( \sum_{j=0}^{k}M_j )p, N-q-( \sum_{j=0}^{k}M_j )p, \alpha}$ represents its $1-\alpha$ quantile.
We reject $H_0$ at $5\%$ level of significance if $F_{ ( \sum_{j=0}^{k}M_j )p, N-q-( \sum_{j=0}^{k}M_j )p} > f_{( \sum_{j=0}^{k}M_j )p, N-q-( \sum_{j=0}^{k}M_j ) p, \alpha}$.
If $H_1$: $\boldsymbol\beta=\tilde{\boldsymbol\beta}\neq 0$ is true, then 
\begin{align*}
\dfrac{N-q-( \sum_{j=0}^{k}M_j )p}{ ( \sum_{j=0}^{k}M_j ) p(N-q-1)} \hat{C}_N(\boldsymbol\delta)\sim F_{ ( \sum_{j=0}^{k}M_j ) p, N-q- ( \sum_{j=0}^{k}M_j ) p, \tilde{C}_N(\boldsymbol\delta)},
\end{align*}
i.e. non-centrality $F$ distribution with non-centrality parameter $\tilde{C}_N(\boldsymbol\delta)$, therefore
\begin{equation}\label{N_F_N_q_1_P}
\text{Pr}\left(F_{ ( \sum_{j=0}^{k}M_j ) p, N-q- ( \sum_{j=0}^{k}M_j ) p, \tilde{C}_N(\boldsymbol\delta)} > f_{ ( \sum_{j=0}^{k}M_j ) p, N-q- ( \sum_{j=0}^{k}M_j )p, \alpha}\right)=\text{Power},
\end{equation}
where $F_{ ( \sum_{j=0}^{k}M_j )p,  N-q-( \sum_{j=0}^{k}M_j )p, \tilde{C}_N(\boldsymbol\delta) )}$ represents a random variable that follows a non-centrality $F$ distribution with degrees of freedom $p\sum_{j=0}^{k}M_j$ and $N-q-( \sum_{j=0}^{k}M_j )p$  and non-centrality parameter  $\tilde{C}_N(\boldsymbol\delta)$.
However, \cite{Liao_etal_2016} does not provide any mathematical proofs for the test statistics distribution. In this paper, we suggest two alternative distributions for $\hat{C}_N(\boldsymbol\delta)$ and provide certain mathematical evidences in the Appendix. These distributions can be defined by
 \begin{align*}
\hat{C}_N(\boldsymbol\delta)=N\hat{\boldsymbol\beta}^{\top}\hat{\boldsymbol\Sigma}_{\boldsymbol\beta\boldsymbol\beta}^{-1}\hat{\boldsymbol\beta}\sim T^2_{ ( \sum_{j=0}^{k}M_j ) p, N-1}=\dfrac{ ( \sum_{j=0}^{k}M_j ) p(N-1)}{N- ( \sum_{j=0}^{k}M_j ) p} F_{ ( \sum_{j=0}^{k}M_j ) p, N- ( \sum_{j=0}^{k}M_j ) p}
\end{align*} 
and
 \begin{align*}
\hat{C}_N(\boldsymbol\delta)=N\hat{\boldsymbol\beta}^{\top}\hat{\boldsymbol\Sigma}_{\boldsymbol\beta\boldsymbol\beta}^{-1}\hat{\boldsymbol\beta}\sim T^2_{ ( \sum_{j=0}^{k}M_j ) p, N}=\dfrac{ ( \sum_{j=0}^{k}M_j ) p(N)}{N- ( \sum_{j=0}^{k}M_j ) p +1 } F_{ ( \sum_{j=0}^{k}M_j ) p, N- ( \sum_{j=0}^{k}M_j ) p +1}
\end{align*} 
when $H_0$ is not rejected. The corresponding type-I error rate ($\alpha$) function can be defined by
 \begin{equation}\label{N_F_N_1_alpha}
\text{Pr}\left(F_{ ( \sum_{j=0}^{k}M_j ) p, N- ( \sum_{j=0}^{k}M_j ) p} > f_{ ( \sum_{j=0}^{k}M_j ) p, N- ( \sum_{j=0}^{k}M_j )p, \alpha}\right)=\alpha
\end{equation}
and
 \begin{equation}\label{N_F_N_alpha}
\text{Pr}\left(F_{ ( \sum_{j=0}^{k}M_j ) p, N- ( \sum_{j=0}^{k}M_j ) p +1} > f_{ ( \sum_{j=0}^{k}M_j ) p, N- ( \sum_{j=0}^{k}M_j )p +1, \alpha}\right)=\alpha
\end{equation}
respectively. 
We reject $H_0$ if 
\begin{equation*}
F_{ ( \sum_{j=0}^{k}M_j ) p, N- ( \sum_{j=0}^{k}M_j ) p} > f_{ ( \sum_{j=0}^{k}M_j ) p, N- ( \sum_{j=0}^{k}M_j )p, \alpha}
\end{equation*}
and 
\begin{equation*}
F_{ ( \sum_{j=0}^{k}M_j ) p, N- ( \sum_{j=0}^{k}M_j ) p +1} > f_{ ( \sum_{j=0}^{k}M_j ) p, N- ( \sum_{j=0}^{k}M_j )p +1, \alpha}
\end{equation*}
at $5\%$ level of significance respectively.
Also if $H_1$: $\boldsymbol\beta=\tilde{\boldsymbol\beta}\neq 0$ is true, then the power function can be defined by
 \begin{equation}\label{N_F_N_1_P}
\text{Pr}\left(F_{ ( \sum_{j=0}^{k}M_j ) p, N- ( \sum_{j=0}^{k}M_j ) p, \tilde{C}_N(\boldsymbol\delta)} > f_{ ( \sum_{j=0}^{k}M_j ) p, N- ( \sum_{j=0}^{k}M_j )p, \alpha}\right)=\text{Power}
\end{equation}
and
 \begin{equation}\label{N_F_N_P}
\text{Pr}\left(F_{ ( \sum_{j=0}^{k}M_j ) p, N- ( \sum_{j=0}^{k}M_j ) p +1, \tilde{C}_N(\boldsymbol\delta)} > f_{ ( \sum_{j=0}^{k}M_j ) p, N- ( \sum_{j=0}^{k}M_j )p +1, \alpha}\right)=\text{Power}
\end{equation}
respectively.

\subsection{Power-Based Sample Size Calculation}\label{powersamplesize}
This section proposes and derives the sample size formula based on power. The power-based method is only used if the goal of the study is to perform a hypothesis test \cite{Yan_etal_2020}.
This method requires that the prior information of the standardized proximal effect size of intervention levels (i.e. $\delta_m(d)=\boldsymbol Z_{d}^{\top}\boldsymbol \delta_m$, where $\boldsymbol \delta_m=\boldsymbol\beta_m/\sigma$, for $m=1,\ldots,M$) are known. 
Thus, given the desired power (i.e. Power) and standardized $\boldsymbol\beta$ ($\boldsymbol\delta$) intended to detect, the sample size $N$ can be computed by solving for either equation (\ref{N_chi_P}), (\ref{N_F_N_q_1_P}), (\ref{N_F_N_1_P}) or (\ref{N_F_N_P}), depending on the choice of test statistic. 
For example, assuming a linearly increasing trend until reaching maximum and constant trend afterwards, $\boldsymbol\delta$ can be derived by the standardized initial ($\delta_m(1)$) and average ($\dfrac{1}{D}\sum_{d=1}^{D}\delta_m(d)$) proximal effect sizes and the day reaching maximum ($d_{\text{max}}$).
Therefore, the calculated $N$ is the minimum integer that gives the power not lower than its nominal value.

\subsection{Precision-Based Sample Size Calculation}\label{precisionsamplesize}
MLMRT is the latest experimental design proposed for the mobile health studies to detect the proximal effect of the ``just in time adaptive intervention". It is novel compared other experimental designs, i.e. randomized controlled trial or sequential multiple assignment randomized trial (SMART). 
There are possibilities that the prior information of the proximal effect is not known. 
Conducting a pilot MLMRT study can benefit researchers to access the feasibility and acceptability of the mobile intervention before conducting a full-scale MLMRT study, similar to Almirall et al \cite{Almirall_etal_2012} who suggested pilot studies for SMART.  

Since the standardized proximal effect size information of a pilot MLMRT study is not available. Hence the power-based method is not practically used to calculate sample size.   
In order to achieve certain level of accuracy, the sample size calculation method based on precision (i.e. $\hat{\boldsymbol\beta} - \tilde{\boldsymbol\beta}$) can be used, e.g., \cite{Kelley_etal_2003} or \cite{Maxwell_etal_2008}. 
Yan et al \cite{Yan_etal_2020} propose and derive the precision-based sample size calculation methods for the pilot SMARTs. 
Therefore, in the similar way, this section, we propose and derive a precision-based sample size calculation method for the our MLMRT.

For the precision-based method, we define the formula of coverage probability ($1-\alpha$) in term of sample size $N$ and precision $\hat{\boldsymbol\beta} - \tilde{\boldsymbol\beta}$ for estimator $\hat{\boldsymbol\beta}$, i.e.
\begin{equation}
\text{Pr}\left(  0 < \dfrac{N-q- ( \sum_{j=0}^{k}M_j ) p}{ ( \sum_{j=0}^{k}M_j ) p(N-q-1)}N( \hat{\boldsymbol\beta} - \tilde{\boldsymbol\beta} )^{\top}\hat{\boldsymbol\Sigma}_{\boldsymbol\beta\boldsymbol\beta}^{-1}( \hat{\boldsymbol\beta} - \tilde{\boldsymbol\beta} ) < f_{ ( \sum_{j=0}^{k}M_j ) p, N-q- ( \sum_{j=0}^{k}M_j ) p, \alpha} \right)=1-\alpha
\end{equation}
assuming $N>q+ ( \sum_{j=0}^{k}M_j ) p$.
Hence $N$ can be calculated by solving the following equation,
\begin{equation}\label{NConfInterv}
( \hat{\boldsymbol\beta} - \tilde{\boldsymbol\beta} )^{\top}\hat{\boldsymbol\Sigma}_{\boldsymbol\beta\boldsymbol\beta}^{-1}( \hat{\boldsymbol\beta} - \tilde{\boldsymbol\beta} ) > \dfrac{ ( \sum_{j=0}^{k}M_j ) p(N-q-1)}{N(N-q- ( \sum_{j=0}^{k}M_j ) p)}f_{ ( \sum_{j=0}^{k}M_j ) p, N-q- ( \sum_{j=0}^{k}M_j ) p, \alpha},
\end{equation}
which has the same distribution for the test statistic as equation (\ref{N_F_N_q_1_P}). 
Therefore, in the similar way, we can derive the sample size formulas based on precision that correspond to equations (\ref{N_chi_P}), (\ref{N_F_N_1_P}) and (\ref{N_F_N_P}), are
\begin{equation}\label{N_chi_C}
( \hat{\boldsymbol\beta} - \tilde{\boldsymbol\beta} )^{\top}\hat{\boldsymbol\Sigma}_{\boldsymbol\beta\boldsymbol\beta}^{-1}( \hat{\boldsymbol\beta} - \tilde{\boldsymbol\beta} ) > \dfrac{ 1 }{ N }\chi_{ ( \sum_{j=0}^{k}M_j ) p, \alpha},
\end{equation}
\begin{equation}\label{N_F_N_1_C}
( \hat{\boldsymbol\beta} - \tilde{\boldsymbol\beta} )^{\top}\hat{\boldsymbol\Sigma}_{\boldsymbol\beta\boldsymbol\beta}^{-1}( \hat{\boldsymbol\beta} - \tilde{\boldsymbol\beta} ) > \dfrac{ ( \sum_{j=0}^{k}M_j ) p(N-1)}{N(N- ( \sum_{j=0}^{k}M_j ) p)}f_{ ( \sum_{j=0}^{k}M_j ) p, N- ( \sum_{j=0}^{k}M_j ) p, \alpha}\end{equation}
and
\begin{equation}\label{N_F_N_C}
( \hat{\boldsymbol\beta} - \tilde{\boldsymbol\beta} )^{\top}\hat{\boldsymbol\Sigma}_{\boldsymbol\beta\boldsymbol\beta}^{-1}( \hat{\boldsymbol\beta} - \tilde{\boldsymbol\beta} ) > \dfrac{ ( \sum_{j=0}^{k}M_j ) p(N)}{N(N- ( \sum_{j=0}^{k}M_j ) p + 1)}f_{ ( \sum_{j=0}^{k}M_j ) p, N- ( \sum_{j=0}^{k}M_j ) p + 1, \alpha}
\end{equation}
respectively. Note that $N$ can be computed by the minimum value that gives the coverage probability not lower than its nominal values ($1-\alpha$).

\section{Simulation 
}\label{s:sim}

In this section, we present simulation studies to investigate the performance of the proposed sample size formulas based on power (i.e. (\ref{N_chi_P}), (\ref{N_F_N_q_1_P}) to (\ref{N_F_N_P})) and precision (i.e. (\ref{N_chi_C}) 
to (\ref{N_F_N_C})). 
For each simulation study, we generate $1000$ Monte Carlo data sets, use desired power $80\%$ and Type-I error rate $\alpha=0.05$. 
The performance of each sample size formula is measured by, comparing the difference between the formulae and Monte Carlo estimates of either power or precision,  based on the calculated $N$.

The Monte Carlo data generation steps are given below. These steps include generating the random variables $I_{id}$, $\boldsymbol A_{id}$, $\epsilon_{id}$ and $Y_{id}$ for each participant $i$, $i=1,\ldots,N$, in each day $d$, $d$=$1,\ldots, D$.
\begin{enumerate}
\item[Step 1.] The availability indicator $I_{id}$ can be generated by binomial random variable, i.e. $I_{id}\sim \text{Binomial}(\tau_{d}, 1)$.
\item[Step 2.] The treatments $\boldsymbol A_{id}$=($A_{i1d}$,$\ldots$,$A_{iM_0d}$,$\ldots$,$A_{i( \sum_{j=0}^{k-1}M_j+1 )d}$,$\ldots$,$A_{i( \sum_{j=0}^{k}M_j )d}$)$^{\top}$ can be generated by multinomial random variable, i.e. $\boldsymbol A_{id}\sim$Multinomial(1-$\sum_m\pi_{md}$, $\pi_{1d}$,$\ldots$,$\pi_{( \sum_{j=0}^{k}M_j  )d}$).
\item[Step 3.] The error term $\boldsymbol\epsilon_{i}$ = ($\epsilon_{i1}$,$\ldots$,$\epsilon_{iD}$)$^{\top}$ can be generated by the multivariate normal random variable, i.e. $\boldsymbol\epsilon_{i}\sim$MVN($\boldsymbol 0_{D\times 1}$, COV($\boldsymbol\epsilon_{i}$)), where COV($\boldsymbol\epsilon_{i}$) is the covariance matrix of $\boldsymbol\epsilon_{i}$ with, dimension of $D\times D$, diagonal of $\sigma^2$ and off diagonal of $\rho\sigma^2$.
In this section, we set $\sigma$=1 and $\rho$=0.
\item[Step 4.] The outcome measure $Y_{id}$ can be computed by  $Y_{id}$ = $\boldsymbol X_{id}^{\top}\boldsymbol\theta$ + $\epsilon_{id}$, where $\boldsymbol\theta$=($\boldsymbol\alpha^\top$, $\boldsymbol\beta^\top$)$^\top$. 
We have $\boldsymbol X_{id}^{\top}$=[$\boldsymbol B_{d}^{\top}$, $( A_{i1d} - \pi_{1d} )\boldsymbol Z_{d}^{\top}$, $\ldots$, $( A_{iM_0d} - \pi_{M_0d}  )\boldsymbol Z_{d}^{\top}$, $\ldots$,$( A_{i( \sum_{j=0}^{k-1}M_j+1 )d} - \pi_{( \sum_{j=0}^{k-1}M_j+1 )d} )\boldsymbol Z_{d}^{\top}$, $\ldots$, $( A_{i( \sum_{j=0}^{k}M_j )d} - \pi_{(\sum_{j=0}^{k}M_j )d} )\boldsymbol Z_{d}^{\top}$ ], where
\begin{equation*}
\boldsymbol B_{d}^{\top} = \left( 1, \ldots,  ( d - 1 )^{q-1} \right)
\end{equation*}
and
\begin{equation*}
\boldsymbol Z_{d}^{\top} = \left( 1, \ldots, ( d - 1 )^{p-1} \right),
\end{equation*}
where we have the constant, linear or quadratic trend of $\boldsymbol Z_d$, which correspond to $p$ = 1, 2 or 3 respectively. Alternatively, we also have mixed of linear and constant trend of $\boldsymbol Z_d$, i.e. 
\begin{equation}\label{Zd_lspline28}
\boldsymbol Z_{d}^{\top} = \left( 1, \text{min}\left[28-1, d - 1 \right] \right),
\end{equation}
where $\boldsymbol Z_d$ is increased linearly until reaching the maximum value at, e.g., 28-day, then maintained the maximum value afterwards. 
\end{enumerate}

Tables \ref{Table: T1} - \ref{Table: T4} present a list of sample sizes calculated by the proposed methods, with the corresponding formula-based and Monte Carlo estimated powers  (i.e. Tables \ref{Table: T1} and \ref{Table: T3}) and coverage probabilities (i.e. Tables \ref{Table: T2} and \ref{Table: T4}). 

We consider the linear increasing trend until maximum at the 28$^{\text{th}}$ day (similar to the data analysis result of HeartSteps study \cite{Klasnja_etal_2019}) and constant (maintain maximum due to adaptive learning messaging technique) trend afterwards for the standardized proximal effect size of intervention message levels, which satisfy $\delta_m(d)=\boldsymbol Z_{d}^{\top}\boldsymbol \delta_m$, where $\boldsymbol \delta_m=\boldsymbol\beta_m/\sigma$, for $m=1,\ldots,M$, and $\boldsymbol Z_{d}^{\top}$ is defined by equation (\ref{Zd_lspline28}). 
As DIAMANTE study assumes each individual is available at each decision time point, this simulation study fixes the availability rate at 100\%.
The study durations ($D$) are $180$, $84$, $28$ and $14$ days. Note that the length of study period of DIAMANTE is six months (approximately 180 days).
The number of decision point per day is $1$, since each participant in DIAMANTE receives each of the intervention components once per day.  
We assume an initial standardized proximal effect size ($\delta_m(1)$) of $0.02$, and precision of an initial standardized proximal effect size ($\hat{\delta}_m(1) - \tilde{\delta}_m(1)$) of $0.02$.  
Note that 0.02 is very close to 0. 
We assume the average standardized proximal effect sizes ($\dfrac{1}{D}\sum_{d=1}^{D}\delta_m(d)$) of $0.2$ and $0.1$ (small effect), and precision of the average standardized proximal effect size ($\dfrac{1}{D}\sum_{d=1}^{D}\left[\hat{\delta}_m(d) - \tilde{\delta}_m(d)\right]$) of $0.25$ and $0.15$.  
We consider number of intervention levels ($M$) of $3$ and $4$, which are consistent with the components of Time Window, Feedback Message and Motivational Message.

Table \ref{Table: T1} computes the sample sizes and the corresponding powers estimated by both formula-based and Monte Carlo methods based on the postulated  effect size and desired power of $0.8$, where we consider three and four intervention levels proposed at the beginning of the study. The allocation probability for control is $0.6$ (based on \cite{Liao_etal_2016}) and the probability of each active intervention level is $0.4/3$ and $0.4/4=0.1$ for $M=3$ and $M=4$ respectively. 
Table \ref{Table: T2} calculates sample sizes based on precision and coverage probability. 
Similar to Table \ref{Table: T1}, Table \ref{Table: T3} computes the sample sizes based on effect size and power, where we have two intervention levels proposed at the beginning of the study and another one (for $M=3$) and two (for $M=4$) levels proposed half way through the study. The allocation probability for control is $0.6$ and the initial probability of each intervention level is $0.2$ ($0.4/2$), the probability changes to $0.4/3$ and $0.1$ ($0.4/4$) after the addition of the new levels. 
Similar way to Table \ref{Table: T3}, Table \ref{Table: T4} calculates the sample sizes based on both precision and coverage probability. 

Figure \ref{fig: MvsN} presents a plot of number of intervention levels versus sample sizes for each type of test statistics. For the number of intervention levels, we consider $M=1$,$\ldots$, $10$, which are all proposed at the beginning of the study. Sample sizes are calculated based on power and effect size, with type-I error rate of $0.05$, desired power of $80\%$, study duration ($D$) of 180-day, only one decision time points per day, standardized proximal effect size of $0.2$,  initial standardized proximal effect size of 0.02, linear increasing trend until reaching maximum at  the $28^{\text{th}}$ and constant trend afterwards for the standardized proximal effect size and $100\%$ availability at each time point. We observe that sample sizes increased linearly with number of intervention message levels for each plot.

The results are summarised below. The tables show that the Monte Carlo estimated powers are close to the corresponding formula-based powers while Monte Carlo estimated coverage probabilities are close to the corresponding formula-based coverage probabilities, except for when sample sizes are small, i.e. when $D=180$ and test statistics$\sim$Hotelling's $T^2_{\sum_{j=1}^kM_jp, N-q-1}$. 
This indicates a suspicion for the distribution of test statistics proposed by Liao et al (2016) that is accurate, and it requires further mathematical proofs. 
One example from Table \ref{Table: T1}, given a component with three intervention levels, 180 days study period and test statistic follows the distribution of Hotelling's $T^{2}_{Mp, N}$, at $5\%$ level of significance, detecting the average standardized proximal effect size of $0.1$ for each of the intervention levels at a sample size of $37$ achieves a power of $85\%$. 
Another example from Table \ref{Table: T2}, maintaining a precision of $0.15$ for the average standardized proximal effect size for each of the intervention levels at a sample size of $22$ results a coverage probability of $96\%$. 
The estimated sample sizes increase with the number of intervention levels, and in the opposite direction of effect sizes or margin of errors and length of study period. Given a particular number of intervention levels, the sample size required is smaller when all the message levels are proposed upfront at the beginning of the study compared to when some of them are proposed after the study onset. 

Similar, we re-run the simulation studies above by considering constant trend for the proximal effect of the intervention messages (e.g. an MRT design for the JOOL study \cite{Bidargaddi_etal_2018}). The similar results are obtained, see Tables \ref{Table: TC5} to \ref{Table: TC8} in Appendix C. 

\newpage
\begin{figure}[ht]
\begin{center}
\begin{tabular}{cc}
 \includegraphics[scale=0.25]{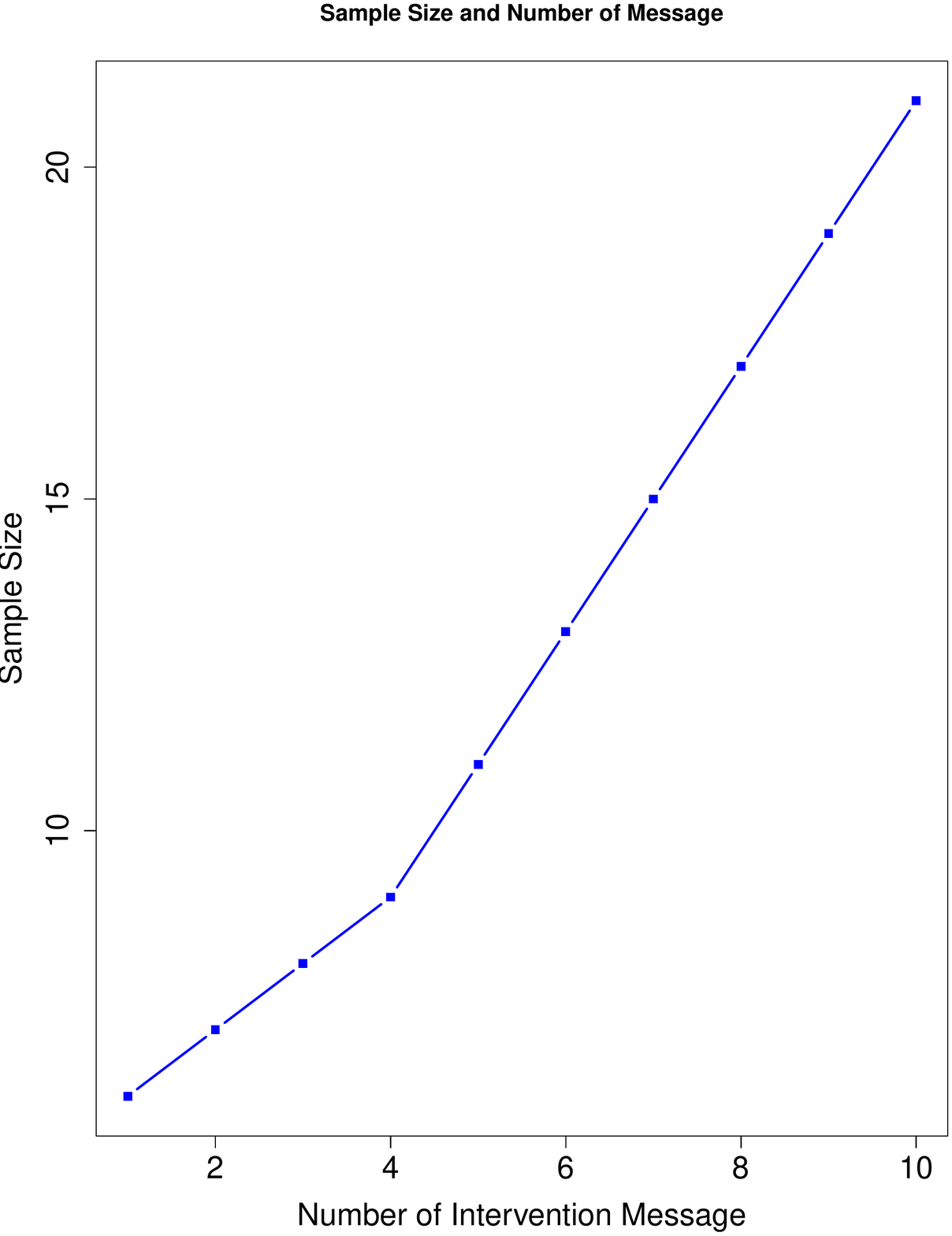} & 
 \includegraphics[scale=0.25]{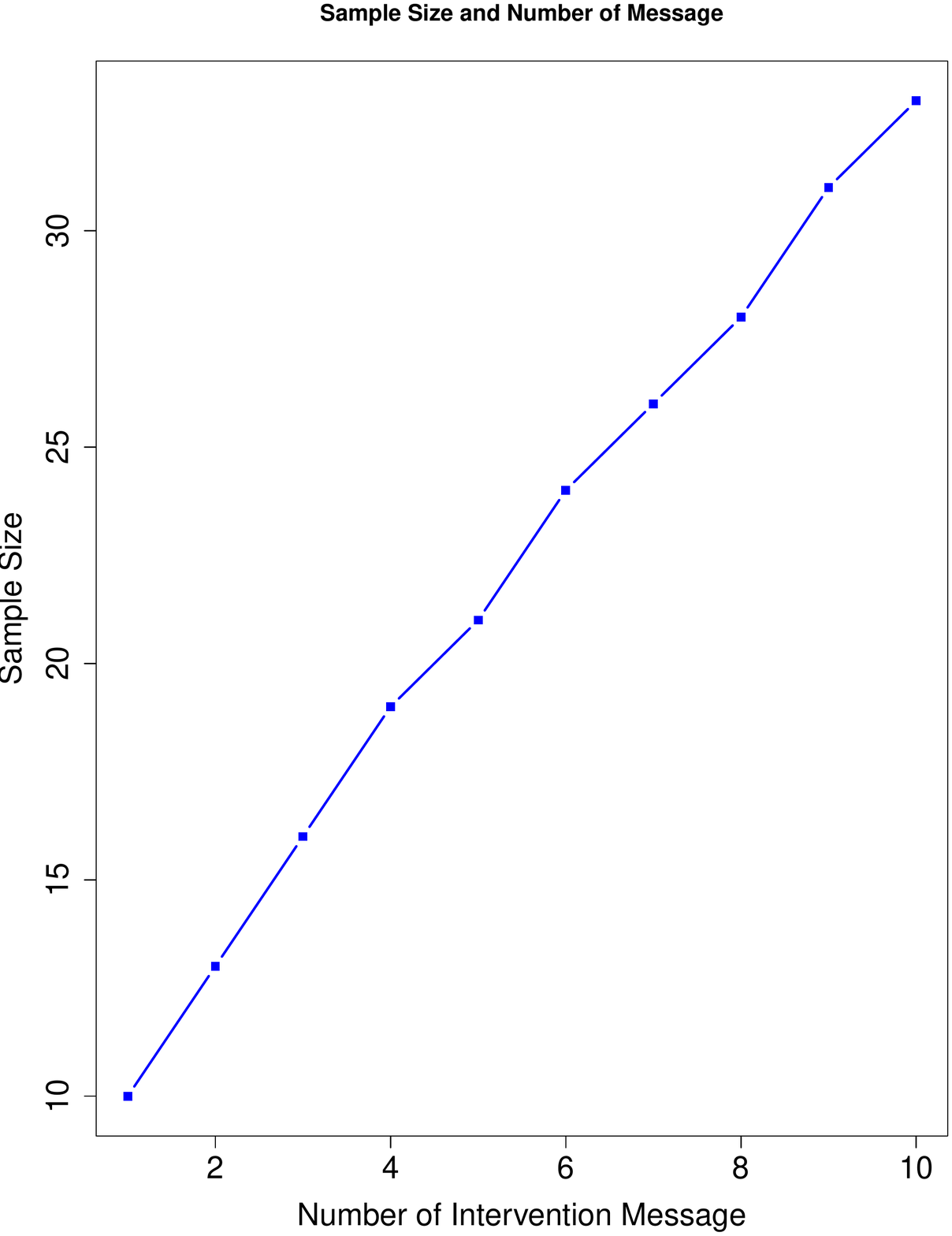} \\
 a) Test Statistics$\sim \chi^2_{Mp}$ & 
 b) Test Statistics$\sim$ Hotelling's $T^2_{Mp, N-q-1}$ \\
  \includegraphics[scale=0.25]{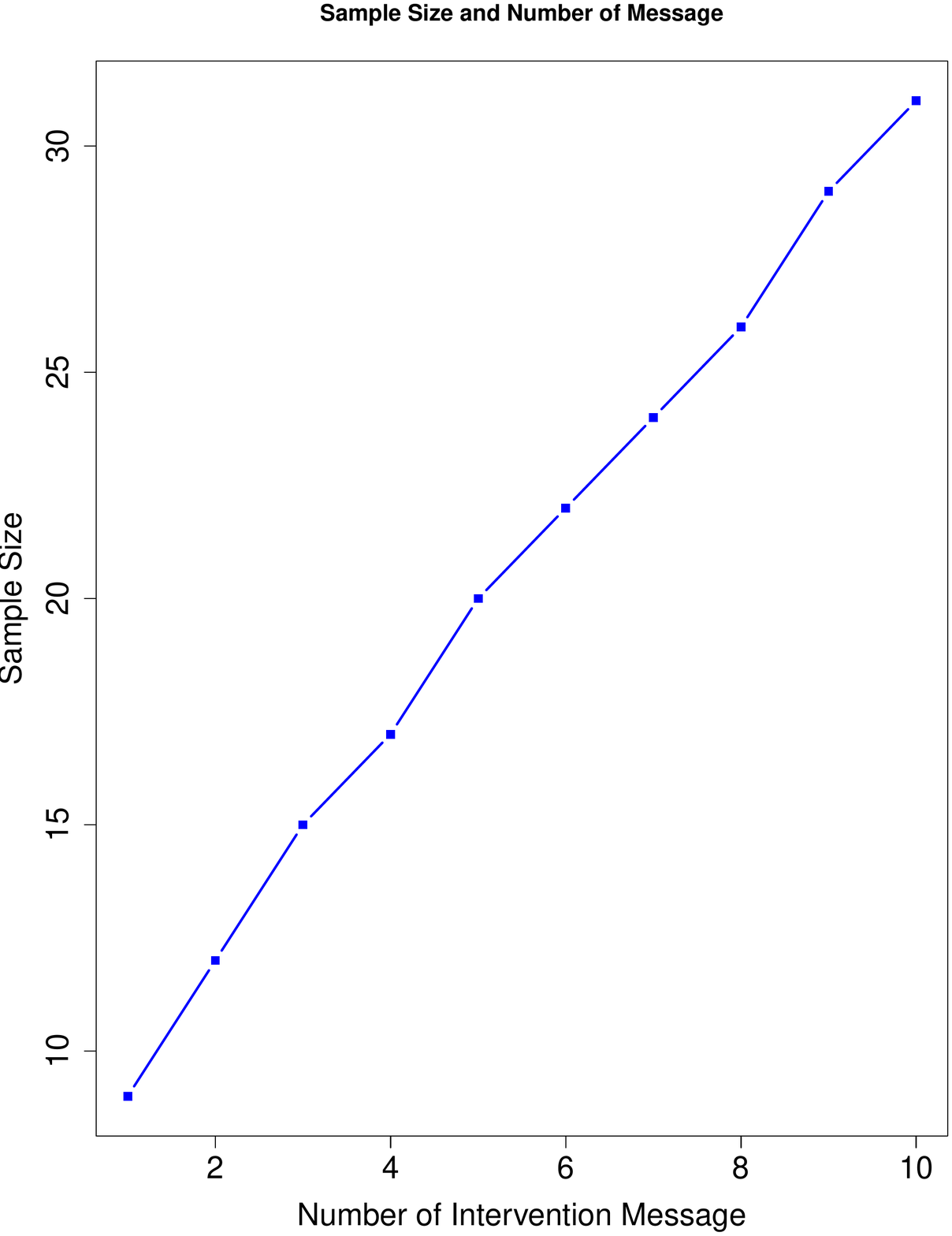} & 
  \includegraphics[scale=0.25]{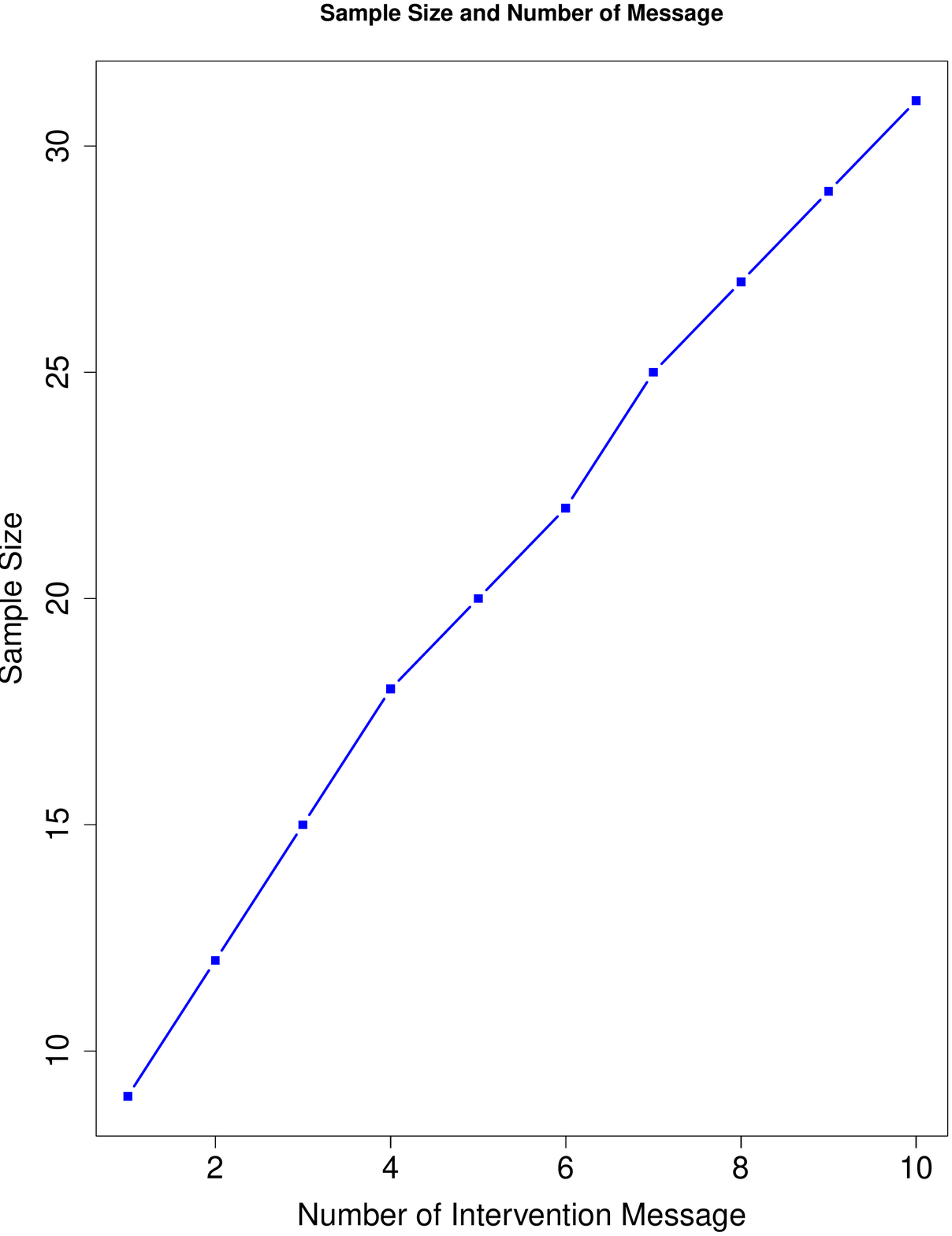} \\
 c) Test Statistics$\sim$ Hotelling's $T^2_{Mp, N}$ & 
 d) Test Statistics$\sim$ Hotelling's $T^2_{Mp, N-1}$ \\
\end{tabular}
\end{center}
\caption{Sample Sizes calculation based on Power when at least one of the standardized proximal effect size of intervention levels satisfies $\delta_m(d)=\boldsymbol Z_{d}^{\top}\boldsymbol \delta_m$, where $\boldsymbol \delta_m=\boldsymbol\beta_m/\sigma$, for $m=1,\ldots,M$. Note that we have $M=1, \ldots, 10$ and $\rho=0$. The significance level is 0.05. The desired power is 0.80. The study duration ($D$) is 180 days. 
The average standardized proximal effect size ($\dfrac{1}{D}\sum_{d=1}^{D}\delta_m(d)$) is 0.2. 
The initial standardized proximal effect size ($\delta_m(1)$) is 0.02. 
Linear increasing trend until reaching maximum at the $28^{\text{th}}$ day and constant trend afterwards for standardized proximal effect size and $100\%$ availability at each time point are assumed. 
}\label{fig: MvsN}
\end{figure}

\newpage
\begin{table}[H]
\caption{Sample sizes calculation based on power (P) when the standardized proximal effect size of intervention levels satisfy $\delta_m(d)=\boldsymbol Z_{d}^{\top}\boldsymbol \delta_m$, where $\boldsymbol \delta_m=\boldsymbol\beta_m/\sigma$, for $m=1,\ldots,M$. 
Note that we have $M=3$, 4, $\sigma=1$ and $\rho=0$. The significance level is 0.05. The desired P is 0.80. ``Duration" is the duration of study ($D$) in days. 
Linear increasing trend until reaching maximum at $28^{\text{th}}$ day and constant trend afterwards for the standardized proximal effect size and $100\%$ availability at each time point are assumed.
The initial standardized proximal effect size is 0.02.
}
\label{Table: T1}
\centering
\begin{tabular}{rllrrrrrrr}
  \hline
 & \multicolumn{3}{c}{} &  \multicolumn{2}{c}{Sample Size } & \multicolumn{2}{c}{Formulated P }  & \multicolumn{2}{c}{Monte Carlo P} \\
 & \multicolumn{3}{c}{} &  \multicolumn{6}{c}{Average standardized proximal effect size}  \\
& Test Statistics & M & Duration & 0.20 & 0.10 & 0.20 & 0.10 & 0.20 & 0.10 \\ 
  \hline
& $\chi^2_{Mp}$ & 3 & 180 &   8 &  31 & 0.83 & 0.81 & 0.83 & 0.82 \\ 
 &  &   &  84 &  16 &  63 & 0.82 & 0.80 & 0.82 & 0.83 \\ 
 &  &   &  28 &  40 & 165 & 0.81 & 0.80 & 0.79 & 0.81 \\ 
 &  &   &  14 &  78 & 326 & 0.80 & 0.80 & 0.82 & 0.79 \\ 
 & Hotelling's $T^2_{Mp, N}$ &   & 180 &  15 &  37 & 0.85 & 0.80 & 0.84 & 0.79 \\ 
 &  &   &  84 &  22 &  70 & 0.80 & 0.81 & 0.80 & 0.80 \\ 
 &  &   &  28 &  46 & 172 & 0.80 & 0.80 & 0.79 & 0.79 \\ 
 &  &   &  14 &  84 & 332 & 0.80 & 0.80 & 0.79 & 0.79 \\ 
 & Hotelling's $T^2_{Mp, N-1}$ &   & 180 &  15 &  38 & 0.82 & 0.81 & 0.78 & 0.80 \\ 
 &  &   &  84 &  23 &  70 & 0.82 & 0.81 & 0.82 & 0.78 \\ 
 &  &   &  28 &  46 & 172 & 0.80 & 0.80 & 0.77 & 0.82 \\ 
 &  &   &  14 &  84 & 332 & 0.80 & 0.80 & 0.78 & 0.82 \\ 
 & Hotelling's $T^2_{Mp, N-q-1}$ &   & 180 &  16 &  38 & 0.82 & 0.81 & 0.75 & 0.77 \\ 
 &  &   &  84 &  23 &  70 & 0.80 & 0.81 & 0.76 & 0.80 \\ 
 &  &   &  28 &  47 & 172 & 0.81 & 0.80 & 0.80 & 0.82 \\ 
 &  &   &  14 &  84 & 332 & 0.80 & 0.80 & 0.78 & 0.81 \\ 
 & $\chi^2_{Mp}$ & 4 & 180 &   9 &  34 & 0.84 & 0.81 & 0.82 & 0.83 \\ 
 &  &   &  84 &  17 &  70 & 0.80 & 0.81 & 0.81 & 0.81 \\ 
 &  &   &  28 &  44 & 182 & 0.81 & 0.80 & 0.80 & 0.81 \\ 
 &  &   &  14 &  86 & 359 & 0.80 & 0.80 & 0.80 & 0.79 \\ 
 & Hotelling's $T^2_{Mp, N}$ &   & 180 &  17 &  42 & 0.82 & 0.81 & 0.80 & 0.80 \\ 
 &  &   &  84 &  26 &  77 & 0.82 & 0.80 & 0.82 & 0.80 \\ 
 &  &   &  28 &  52 & 190 & 0.81 & 0.80 & 0.80 & 0.80 \\ 
 &  &   &  14 &  94 & 367 & 0.81 & 0.80 & 0.79 & 0.80 \\ 
 & Hotelling's $T^2_{Mp, N-1}$ &   & 180 &  18 &  42 & 0.84 & 0.80 & 0.81 & 0.80 \\ 
 &  &   &  84 &  26 &  78 & 0.82 & 0.81 & 0.79 & 0.79 \\ 
 &  &   &  28 &  52 & 190 & 0.81 & 0.80 & 0.78 & 0.81 \\ 
 &  &   &  14 &  94 & 367 & 0.80 & 0.80 & 0.79 & 0.79 \\ 
 & Hotelling's $T^2_{Mp , N-q-1}$ &   & 180 &  19 &  43 & 0.84 & 0.81 & 0.80 & 0.80 \\ 
 &  &   &  84 &  27 &  78 & 0.82 & 0.80 & 0.76 & 0.79 \\ 
 &  &   &  28 &  52 & 190 & 0.80 & 0.80 & 0.78 & 0.81 \\ 
 &  &   &  14 &  94 & 367 & 0.80 & 0.80 & 0.81 & 0.79 \\    \hline
    \hline
\end{tabular}
\end{table}

\begin{table}[H]
\caption{Sample sizes calculation based on coverage probability (CP) when the standardized proximal effect size of intervention levels satisfy $\delta_m(d)=\boldsymbol Z_{d}^{\top}\boldsymbol \delta_m$, where $\boldsymbol \delta_m=\boldsymbol\beta_m/\sigma$, for $m=1,\ldots,M$. Note that we have $M=3$, 4, $\sigma=1$ and $\rho=0$. 
The desired CP is $95\%$.
 ``Duration" is the duration of study ($D$) in days. 
 Linear increasing trend until reaching maximum at the $28^{\text{th}}$ day and constant trend afterwards for the standardized proximal effect size and $100\%$ availability at each time point are assumed.
 Precision of initial standardized proximal effect size is 0.02.
}
\label{Table: T2}
\centering
\begin{tabular}{rllrrrrrrr}
  \hline
  & \multicolumn{3}{c}{} & \multicolumn{2}{c}{ Sample Size } & \multicolumn{2}{c}{ Formulated CP }  & \multicolumn{2}{c}{ Monte Carlo CP } \\
 & \multicolumn{3}{c}{} &  \multicolumn{6}{c}{Precision of average standardized proximal effect size}  \\
 & Test Statistics & M & Duration & 0.25 & 0.15 & 0.25 & 0.15 & 0.25 & 0.15 \\ 
  \hline
& $\chi^2_{Mp}$ & 3 & 180 &   7 &  13 & 1.00 & 0.96 & 1.00 & 0.96 \\ 
 &  &   &  84 &  10 &  26 & 0.97 & 0.96 & 0.95 & 0.95 \\ 
 &  &   &  28 &  24 &  66 & 0.96 & 0.95 & 0.95 & 0.95 \\ 
 &  &   &  14 &  46 & 130 & 0.95 & 0.95 & 0.94 & 0.95 \\ 
 & Hotelling's $T^2_{Mp, N}$ &   & 180 &  13 &  22 & 0.96 & 0.96 & 0.95 & 0.97 \\ 
 &  &   &  84 &  18 &  35 & 0.95 & 0.95 & 0.96 & 0.95 \\ 
 &  &   &  28 &  32 &  75 & 0.95 & 0.95 & 0.96 & 0.96 \\ 
 &  &   &  14 &  55 & 139 & 0.95 & 0.95 & 0.95 & 0.95 \\ 
 & Hotelling's $T^2_{Mp, N-1}$ &   & 180 &  14 &  22 & 0.96 & 0.95 & 0.97 & 0.97 \\ 
 &  &   &  84 &  19 &  35 & 0.96 & 0.95 & 0.96 & 0.95 \\ 
 &  &   &  28 &  33 &  75 & 0.96 & 0.95 & 0.97 & 0.96 \\ 
 &  &   &  14 &  55 & 139 & 0.95 & 0.95 & 0.94 & 0.94 \\ 
 & Hotelling's $T^2_{Mp, N-q-1}$ &   & 180 &  15 &  23 & 0.96 & 0.96 & 0.99 & 0.97 \\ 
 &  &   &  84 &  20 &  36 & 0.96 & 0.96 & 0.98 & 0.96 \\ 
 &  &   &  28 &  33 &  76 & 0.95 & 0.95 & 0.97 & 0.95 \\ 
 &  &   &  14 &  55 & 139 & 0.95 & 0.95 & 0.95 & 0.95 \\ 
& $\chi^2_{Mp}$ & 4 & 180 &   9 &  16 & 1.00 & 0.96 & 1.00 & 0.96 \\ 
&  &   &  84 &  12 &  32 & 0.97 & 0.96 & 0.96 & 0.94 \\ 
&  &   &  28 &  29 &  81 & 0.96 & 0.95 & 0.94 & 0.95 \\ 
&  &   &  14 &  56 & 160 & 0.95 & 0.95 & 0.93 & 0.95 \\ 
& Hotelling's $T^2_{Mp, N}$ &  & 180 &  17 &  27 & 0.97 & 0.96 & 0.95 & 0.96 \\ 
&  &   &  84 &  23 &  43 & 0.96 & 0.95 & 0.96 & 0.95 \\ 
&  &   &  28 &  40 &  93 & 0.95 & 0.95 & 0.95 & 0.94 \\ 
&  &   &  14 &  68 & 171 & 0.95 & 0.95 & 0.96 & 0.95 \\ 
& Hotelling's $T^2_{Mp, N-1}$ &  & 180 &  17 &  27 & 0.96 & 0.95 & 0.96 & 0.96 \\ 
 &  &   &  84 &  23 &  43 & 0.96 & 0.95 & 0.96 & 0.96 \\ 
  &  &   &  28 &  40 &  93 & 0.95 & 0.95 & 0.94 & 0.95 \\ 
  &  &   &  14 &  68 & 171 & 0.95 & 0.95 & 0.94 & 0.94 \\ 
  & Hotelling's $T^2_{Mp, N-q-1}$ &  & 180 &  19 &  28 & 0.97 & 0.95 & 0.98 & 0.97 \\ 
  &  &   &  84 &  24 &  44 & 0.95 & 0.95 & 0.97 & 0.95 \\ 
  &  &   &  28 &  41 &  93 & 0.95 & 0.95 & 0.96 & 0.96 \\ 
  &  &   &  14 &  68 & 172 & 0.95 & 0.95 & 0.96 & 0.96 \\      
  \hline
\end{tabular}
\end{table}

\begin{table}[H]
\caption{Sample sizes calculation based on power (P) when the standardized proximal effect size of intervention levels satisfy $\delta_m(d)=\boldsymbol Z_{d}^{\top}\boldsymbol \delta_m$, where $\boldsymbol \delta_m=\boldsymbol\beta_m/\sigma$, for $m=1,\ldots,M_0$,$\ldots$,$\sum_{j=0}^{k}M_j$. Note that we have $k=1$, $M_0=2$ and $M_1=1$ ($M=3$), $M_0=2$ and $M_1=2$ ($M=4$), $\sigma=1$ and $\rho=0$, where $d_0=1$ and $d_1$ is the half way through ``Duration", which is the duration of study ($D$) in days, e.g., if $D=28$ then $d_1=15$. The significance level is 0.05. The desired power is 0.80. 
Linear increasing trend until reaching maximum at the $28^{\text{th}}$ day and constant trend afterwards for the standardized proximal effect size and $100\%$ availability at each decision time point are assumed.
The initial standardized proximal effect size is 0.02.
}
\label{Table: T3}
\centering
\begin{tabular}{rllrrrrrrr}
  \hline
   & \multicolumn{3}{c}{} &  \multicolumn{2}{c}{ Sample Size } & \multicolumn{2}{c}{ Formulated P }  & \multicolumn{2}{c}{ Monte Carlo P } \\
 & \multicolumn{3}{c}{} &  \multicolumn{6}{c}{ Average standardized proximal effect size }  \\
& Test Statistics & M & Duration & 0.20 & 0.10 & 0.20 & 0.10 & 0.20 & 0.10 \\  
  \hline
& $\chi^2_{Mp}$ & 3 & 180 &   8 &  31 & 0.82 & 0.81 & 0.81 & 0.80 \\ 
 &  &   &  84 &  16 &  64 & 0.81 & 0.80 & 0.84 & 0.80 \\ 
 &  &   &  28 &  44 & 183 & 0.80 & 0.80 & 0.79 & 0.80 \\ 
 &  &   &  14 &  86 & 360 & 0.80 & 0.80 & 0.79 & 0.80 \\ 
 & Hotelling's $T^2_{Mp, N}$ &   & 180 &  15 &  38 & 0.84 & 0.81 & 0.82 & 0.80 \\ 
 &  &   &  84 &  23 &  70 & 0.83 & 0.80 & 0.82 & 0.80 \\ 
 &  &   &  28 &  51 & 189 & 0.81 & 0.80 & 0.80 & 0.78 \\ 
 &  &   &  14 &  93 & 366 & 0.81 & 0.80 & 0.79 & 0.79 \\ 
 & Hotelling's $T^2_{Mp, N-1}$ &   & 180 &  15 &  38 & 0.82 & 0.81 & 0.78 & 0.81 \\ 
 &  &   &  84 &  23 &  71 & 0.82 & 0.81 & 0.78 & 0.82 \\ 
 &  &   &  28 &  51 & 189 & 0.81 & 0.80 & 0.80 & 0.83 \\ 
 &  &   &  14 &  93 & 366 & 0.80 & 0.80 & 0.80 & 0.79 \\ 
 & Hotelling's $T^2_{Mp, N-q-1}$ &   & 180 &  16 &  38 & 0.81 & 0.81 & 0.73 & 0.77 \\ 
 &  &   &  84 &  24 &  71 & 0.83 & 0.81 & 0.77 & 0.78 \\ 
 &  &   &  28 &  51 & 189 & 0.81 & 0.80 & 0.78 & 0.79 \\ 
 &  &   &  14 &  93 & 366 & 0.80 & 0.80 & 0.76 & 0.77 \\ 
 & $\chi^2_{Mp}$ & 4 & 180 &   9 &  34 & 0.83 & 0.80 & 0.81 & 0.80 \\ 
 &  &   &  84 &  18 &  71 & 0.82 & 0.80 & 0.80 & 0.79 \\ 
 &  &   &  28 &  52 & 214 & 0.81 & 0.80 & 0.80 & 0.81 \\ 
 &  &   &  14 & 101 & 421 & 0.80 & 0.80 & 0.82 & 0.80 \\ 
 & Hotelling's $T^2_{Mp, N}$ &   & 180 &  17 &  42 & 0.81 & 0.80 & 0.81 & 0.80 \\ 
 &  &   &  84 &  26 &  79 & 0.81 & 0.80 & 0.79 & 0.80 \\ 
 &  &   &  28 &  60 & 222 & 0.81 & 0.80 & 0.79 & 0.80 \\ 
 &  &   &  14 & 109 & 429 & 0.80 & 0.80 & 0.78 & 0.79 \\ 
 & Hotelling's $T^2_{Mp, N-1}$ &   & 180 &  18 &  43 & 0.84 & 0.81 & 0.79 & 0.79 \\ 
 &  &   &  84 &  26 &  79 & 0.80 & 0.80 & 0.76 & 0.79 \\ 
 &  &   &  28 &  60 & 222 & 0.81 & 0.80 & 0.80 & 0.78 \\ 
 &  &   &  14 & 109 & 429 & 0.80 & 0.80 & 0.79 & 0.78 \\ 
 & Hotelling's $T^2_{Mp, N-q-1}$ &   & 180 &  19 &  43 & 0.83 & 0.81 & 0.76 & 0.79 \\ 
 &  &   &  84 &  27 &  80 & 0.81 & 0.81 & 0.78 & 0.77 \\ 
 &  &   &  28 &  60 & 222 & 0.80 & 0.80 & 0.80 & 0.77 \\ 
 &  &   &  14 & 109 & 429 & 0.80 & 0.80 & 0.81 & 0.79 \\     \hline
    \hline
\end{tabular}

\end{table}

\begin{table}[H]
\caption{Sample sizes calculation based on coverage probability when the standardized proximal effect size of intervention levels satisfiy $\delta_m(d)=\boldsymbol Z_{d}^{\top}\boldsymbol \delta_m$, where $\boldsymbol \delta_m=\boldsymbol\beta_m/\sigma$, for $m=1,\ldots,M_0$,$\ldots$,$\sum_{j=0}^{k}M_j$. Note that we have $k=1$, $M_0=2$ and $M_1=1$ ($M=3$), $M_0=2$ and $M_2=1$ ($M=4$), $\sigma=1$ and $\rho=0$, where $d_0=1$ and $d_1$ is the half way through ``Duration", which is the duration of study ($D$) in days, e.g., if $D=28$ then $d_1=15$.  
The desired CP is $95\%$.
Linear increasing trend until reaching maximum at $28^{\text{th}}$ day and constant trend afterwards for standardized proximal effect size and $100\%$ availability at each time point are assumed.
Precision of initial standardized proximal effect size is 0.02.
}
\label{Table: T4}
\centering
\begin{tabular}{rllrrrrrrr}
  \hline
  & \multicolumn{3}{c}{} & \multicolumn{2}{c}{ Sample Size } & \multicolumn{2}{c}{ Formulated CP } & \multicolumn{2}{c}{ Monte Carlo CP } \\
 & \multicolumn{3}{c}{} &  \multicolumn{6}{c}{Precision of average standardized proximal effect size}  \\
& Test Statistics & M & Duration & 0.25 & 0.15 & 0.25 & 0.15 & 0.25 & 0.15 \\ 
  \hline
& $\chi^2_{Mp}$ & 3 & 180 &   7 &  13 & 1.00 & 0.96 & 0.99 & 0.95 \\ 
   &  &   &  84 &  10 &  26 & 0.97 & 0.95 & 0.97 & 0.95 \\ 
   &  &   &  28 &  26 &  73 & 0.95 & 0.95 & 0.94 & 0.94 \\ 
   &  &   &  14 &  51 & 144 & 0.95 & 0.95 & 0.96 & 0.95 \\ 
   & Hotelling's $T^2_{Mp, N}$ &   & 180 &  13 &  22 & 0.96 & 0.96 & 0.95 & 0.96 \\ 
   &  &   &  84 &  18 &  35 & 0.95 & 0.95 & 0.95 & 0.96 \\ 
   &  &   &  28 &  35 &  83 & 0.95 & 0.95 & 0.95 & 0.95 \\ 
   &  &   &  14 &  60 & 153 & 0.95 & 0.95 & 0.95 & 0.96 \\ 
   & Hotelling's $T^2_{Mp, N-1}$ &   & 180 &  14 &  22 & 0.96 & 0.95 & 0.97 & 0.96 \\ 
  &  &   &  84 &  19 &  35 & 0.96 & 0.95 & 0.98 & 0.95 \\ 
   &  &   &  28 &  35 &  83 & 0.95 & 0.95 & 0.96 & 0.96 \\ 
   &  &   &  14 &  60 & 153 & 0.95 & 0.95 & 0.94 & 0.96 \\ 
  & Hotelling's $T^2_{Mp, N-q-1}$ &   & 180 &  15 &  23 & 0.96 & 0.96 & 0.98 & 0.97 \\ 
   &  &   &  84 &  20 &  36 & 0.96 & 0.95 & 0.97 & 0.96 \\ 
   &  &   &  28 &  36 &  83 & 0.95 & 0.95 & 0.96 & 0.96 \\ 
   &  &   &  14 &  60 & 153 & 0.95 & 0.95 & 0.95 & 0.95 \\ 
   & $\chi^2_{Mp}$ & 4 & 180 &   9 &  16 & 1.00 & 0.96 & 1.00 & 0.95 \\ 
   &  &   &  84 &  12 &  33 & 0.96 & 0.96 & 0.95 & 0.96 \\ 
   &  &   &  28 &  34 &  96 & 0.95 & 0.95 & 0.94 & 0.95 \\ 
   &  &   &  14 &  66 & 188 & 0.95 & 0.95 & 0.95 & 0.95 \\ 
   & Hotelling's $T^2_{Mp, N}$ &   & 180 &  17 &  27 & 0.97 & 0.95 & 0.96 & 0.97 \\ 
   &  &   &  84 &  23 &  44 & 0.96 & 0.95 & 0.96 & 0.97 \\ 
   &  &   &  28 &  46 & 108 & 0.96 & 0.95 & 0.96 & 0.96 \\ 
   &  &   &  14 &  78 & 200 & 0.95 & 0.95 & 0.95 & 0.96 \\ 
   & Hotelling's $T^2_{Mp, N-1}$ &   & 180 &  17 &  28 & 0.95 & 0.96 & 0.95 & 0.96 \\ 
   &  &   &  84 &  23 &  44 & 0.95 & 0.95 & 0.96 & 0.96 \\ 
  &  &   &  28 &  46 & 108 & 0.95 & 0.95 & 0.96 & 0.94 \\ 
   &  &   &  14 &  78 & 200 & 0.95 & 0.95 & 0.96 & 0.95 \\ 
   & Hotelling's $T^2_{Mp, N-q-1}$ &   & 180 &  19 &  28 & 0.97 & 0.95 & 0.98 & 0.96 \\ 
   &  &   &  84 &  24 &  45 & 0.95 & 0.95 & 0.97 & 0.95 \\ 
   &  &   &  28 &  46 & 108 & 0.95 & 0.95 & 0.95 & 0.96 \\ 
   &  &   &  14 &  78 & 200 & 0.95 & 0.95 & 0.95 & 0.94 \\       
   \hline
\end{tabular}
\end{table}  

\section{\texttt{R} implementation}\label{s: rimp}

\subsection{Implementation in \texttt{R} function}\label{s:demo}
\label{s:demo}
In this section, we demonstrate the implementation of the \texttt{R} function \texttt{SampleSize\_MLMRT} for sample size calculations via a simulation study. This function is available in  
the GitHub link \url{https://github.com/Kenny-Jing-Xu/MLMRT-SS/blob/master/SampleSizeMLMRT.R}. 

The inputs of the function \\
\begin{tcolorbox}[boxrule=0pt]
\begin{alltt}
SampleSize\_MLMRT(days, occ\_per\_day, aa.day.aa, prob, 
        beta\_shape, beta\_mean, beta\_initial, beta\_quadratic_max, 
        tau\_shape, tau\_mean, tau\_initial, tau\_quadratic\_max, 
        sigma, pow, sigLev, method, test, result, SS) 
\end{alltt}
\end{tcolorbox}
are defined in the following.
\begin{itemize}
\item \texttt{days}: the number of days during the study period.
\item \texttt{occ\_per\_day}: the number of decision time point per day during the study period. Note it is fixed at \texttt{occ\_per\_day=1} for DIAMANTE study.
\item \texttt{aa.day.aa}: the day of each proposed interventional level.
\item \texttt{prob}: the allocation probability matrix of the levels, with dimension of D by (M+1), where the first column are the allocation probabilities of control level.
\item \texttt{beta\_shape}: the shape of $\beta(d)$ in terms of $d$, i.e. ``constant'', ``linear'', ``linear and constant'', or ``quadratic".
\item \texttt{beta\_mean}: the average of standardized proximal effect size of each interventional level over the study period.
\item \texttt{beta\_initial}: the initial standardized proximal effect size of each interventional level.
\item \texttt{beta\_quadratic\_max}: the first value of $d$ that gives the maximum value of $\beta(d)$ if the shape is ``quadratic'' or ``linear and constant''.
\item \texttt{tau\_shape}: the shape of $\tau_{d}$ in term of $d$, i.e. ``constant", ``linear", ``linear and constant", or ``quadratic".
\item \texttt{tau\_mean}: the average of $\tau_{d}$ over the study period.
\item \texttt{tau\_initial}: the initial $\tau_{d}$ .
\item \texttt{tau\_quadratic\_max}: the value of $d$ that gives the maximum value of $\tau_{d}$ if the shape is quadratic
\item \texttt{sigma}: $\sigma$ or the standard deviation of the error term $\epsilon$
\item \texttt{pow}: the power
\item \texttt{sigLev}: the type-I error rate
\item \texttt{method}: the method of sample size calculation, based on either power, i.e. ``power" or confidence interval, i.e. ``confidence interval"
\item \texttt{test}: the test statistics, based on Chi-squared distribution, i.e. ``chi", Hotelling's T-squared distribution with $N-q-1$ degrees of freedom, i.e. ``hotelling N-q-1",  Hotelling's T-squared distribution with $N-1$ degrees of freedom, i.e. ``hotelling N-1", or Hotelling's T-squared distribution with $N$ degrees of freedom, i.e. ``hotelling N"    
\item \texttt{result}: the chosen calculated result, i.e ``choice\_sample\_size", ``choice\_power" or ``choice\_coverage\_probability"
\item \texttt{SS}: the specified sample size if the result of either power or coverage probability is chosen 
\end{itemize}

The possible outputs are summarized below:
\begin{itemize}
\item \texttt{N}, the estimated sample size;
\item \texttt{P}, the estimated power;
\item \texttt{CP}, the estimated coverage probability;
\item \texttt{d}, the consistent estimated $\boldsymbol\beta$;
\item \texttt{Sig\_bet\_inv}, the inverse of the asymptotic covariance estimated for $\boldsymbol\beta$;
\end{itemize}

Here we present the \ts{R} code for the sample size  calculation corresponding to 
\begin{itemize}
\item Test statistics=Hotelling's $T^2_{Mp, N}$,
\item $M=4$,
\item $D=180$,
\item Average proximal standardised effect=$0.2$.
\end{itemize} 
\setlist{nolistsep}
from Table \ref{Table: T3} in Section \ref{s:sim}.
\newpage
\begin{tcolorbox}[boxrule=0pt]
\begin{alltt}
#One decision time point per day during the study period 
occ_per_day=1 

# The regression coefficient of B(d)
alpha1=c(1, 1) 

#The number of parameter of each beta, i.e. p
pb=2

# The study period of 180-day
days=180

# Two interventional levels proposed at the beginning and 
#another two proposed in the middle of the trial.
aa.each=c(2, 2)

# The first day, the middle day and the last day of the study period
aa.day=c(1, ( floor(days/2) + 1 ), days)

# The number of days when the interventional levels
# are proposed in the trial
aa.freq=length(aa.day)-1

# The days that each of the interventional levels
# are added to the trial, e.g., c(1,1,91,91)  
aa.day.aa=rep( ( aa.day )[1:aa.freq], aa.each )

# The allocation probabilities for control level (e.g., 0.6) and 
# interventional levels (e.g., 0.4)  
prob.con=0.6
prob.arm=0.4

# The allocation probability matrix, with dimension 180 by 5 
# The first column are the allocation probability of control level 0.6. 
# Other columns are the probabilities of interventional levels.
# Rows 1-90 are c(0.6,0.2,0.2,0,0) and rows 91-180 are c(0.6,0.1,0.1,0.1,0.1).
prob = matrix( 0, days, ( 1 + sum(aa.each) ) )
prob[,1]=rep(prob.con, days)
for( j in 1:( length( aa.day ) - 1 ) )
\{
  prob[(aa.day[j + 0] + 0):(aa.day[j + 1] - 1), 2:(1 + cumsum(aa.each)[j])] = 
  prob.arm/(cumsum(aa.each)[j])
\}
prob[(aa.day[length(aa.day)]), 2:(1 + sum(aa.each))] = prob.arm/(sum(aa.each))
prob
\end{alltt}
\end{tcolorbox}

\begin{tcolorbox}[boxrule=0pt]
\begin{alltt}
# Linear until 28-day then constant in term of d for the beta shape
beta_shape= "linear and constant"
# Average proximal standardised effect size for each interventional level
beta_mean=rep(0.2, sum( aa.each ) ) 
# Initial proximal standardised effect size for each interventional level
beta_initial=rep(0.02, sum( aa.each ) )
# Maximum proximal standardised effect day for each interventional level
beta_quadratic_max=aa.day.aa-1+28

# Constant shape for availability proportion
tau_shape="constant" 
# Average availability proportion
tau_mean=1
# Initial availability proportion
tau_initial=1
# Maximum availability day
tau_quadratic_max=28

# The power 0.8 and type-I error rate 0.05
pow=0.8 
sigLev=0.05

# The standard deviation and correlation coefficient of the error term
sigma=1
rho=0

# The sample size calculated  by power and test statistics with distribution 
# of Hotelling's T-squared with degrees of freedom N
method="power"
test = "hotelling N"
result = "choice_sample_size"

MRTN=SampleSize_MLMRT(days=days, occ_per_day=occ_per_day, 
                     aa.day.aa = aa.day.aa, 
                     prob=prob, 
                     beta_shape=beta_shape, beta_mean=beta_mean, 
                     beta_initial=beta_initial, 
                     beta_quadratic_max=beta_quadratic_max, 
                     tau_shape=tau_shape, tau_mean=tau_mean, 
                     tau_initial=tau_initial, 
                     tau_quadratic_max=tau_quadratic_max, 
                     sigma=sigma, pow=pow, sigLev=sigLev, 
                     method=method, test=test, result=result)
\end{alltt}
\end{tcolorbox}
                     
\begin{tcolorbox}[boxrule=0pt]
\begin{alltt}
# The calculated sample size
N=MRTN$N
\end{alltt}
\end{tcolorbox}

The above R codes calculate sample size $17$.  The formulated power achieved under this number of participants ($0.81$) can be calculated by substituting the inputs of
\texttt{result="choice\_power"}
and
 \texttt{SS=17} 
in the \texttt{R} function \texttt{SampleSize\_MLMRT}.   

\subsection{Implementation using \texttt{R} shiny}\label{s: Rshiny}
Similar to the online sample size calculator (MRT-SS) for the micro-randomized trial created by Seewald and Liao (2016) \cite{Seewald_Liao_2016} using \texttt{R} shiny, we create our sample size calculator named ``MLMRT-SS'' for the proposed multi-level micro-randomised trial.
The web link of this application is \url{https://kennyxu.shinyapps.io/mlmrt_shinyapps/}. We explore each of its components briefly using Figure \ref{fig: shiny} based on the same example of Section \ref{s:demo}. 

Figure \ref{fig: shiny} describes a trial that has study period $180$ days with only $1$ decision time point per day. For a particular component, only $2$ intervention levels are proposed before the first day of study, and another $2$ levels are added in the half-way through the study, i.e. at $90$ days.
 At each decision time point, a participant is randomly allocated to either the control level or one of the active intervention levels. 
 The total randomization probabilities of the intervention level is $0.4$. For example, the randomization probability for each level is $0.2$ ($0.4/2$) for the first half of the study period, while the randomization probability for each level is $0.1$ ($0.4/4$) for the second half of the study period when two more levels are added.  
 The availability of each participant is expected $100\%$ at all the decision time points. 
The sample size calculation method is based on power. The test statistic is assumed to follow a Hotelling's $T^2$ distribution with denominator degrees of freedom $N$.
For the proximal effect, it is assumed to follow a trend of increasing linearly from day-1 to day-28 and then maintaining constantly at its maximum value. The initial and average values of the standardized proximal effect size for either intervention level are $0.02$ and $0.2$ respectively.
In the `Result' section, the sample size is calculated as the final output under a nominal $80\%$ power and a level of significance $0.05$. The output is presented by ``The required sample size is 17 to attain 80\% power when the significance level is 0.05.'' after the `Get Result' botton is clicked. 
Alternatively, if `Power' is selected in the `Result' section, and $17$ is specified in the `Number of Participants' cell, the output is going to be ``The sample size 17 gives 81\% power when the significance level is 0.05''.

\newpage
\begin{figure}[H]
\begin{center}
\begin{tabular}{cc}
 \includegraphics[height=5cm, width=8cm]{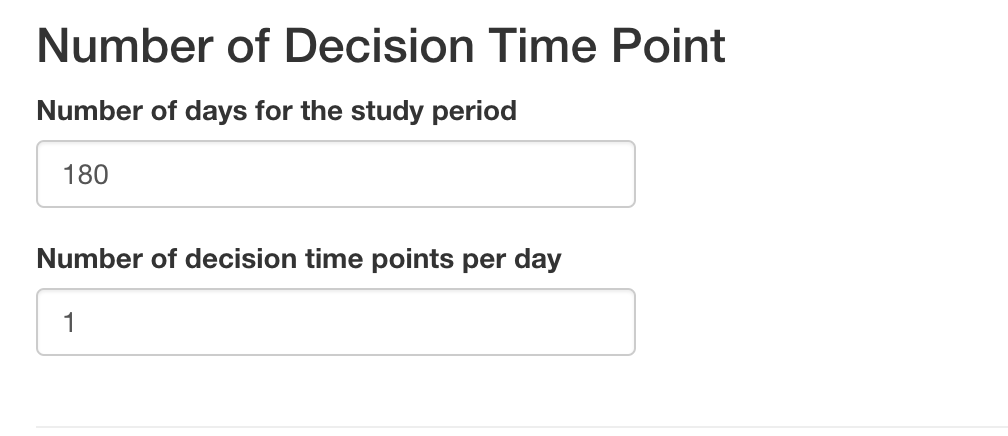} & \includegraphics[height=5cm, width=8cm]{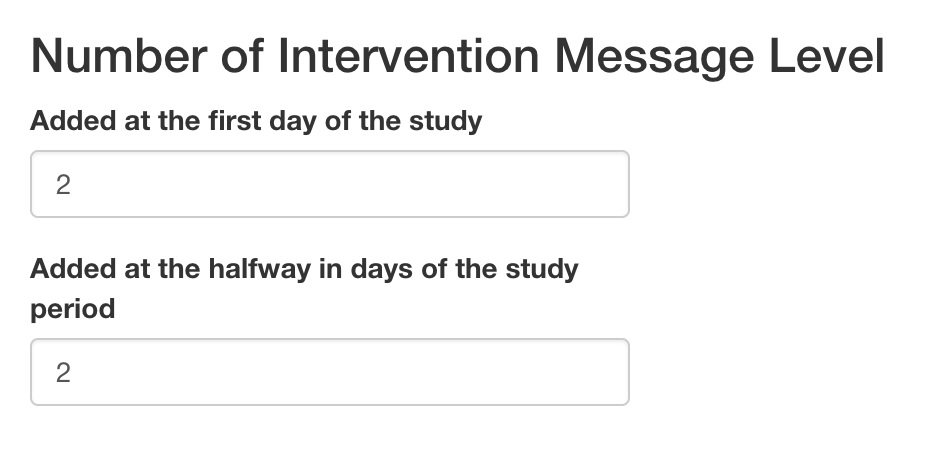} \\
 a) Number of decision time point & b) Number of intervention message level \\
  \includegraphics[height=5cm, width=8cm]{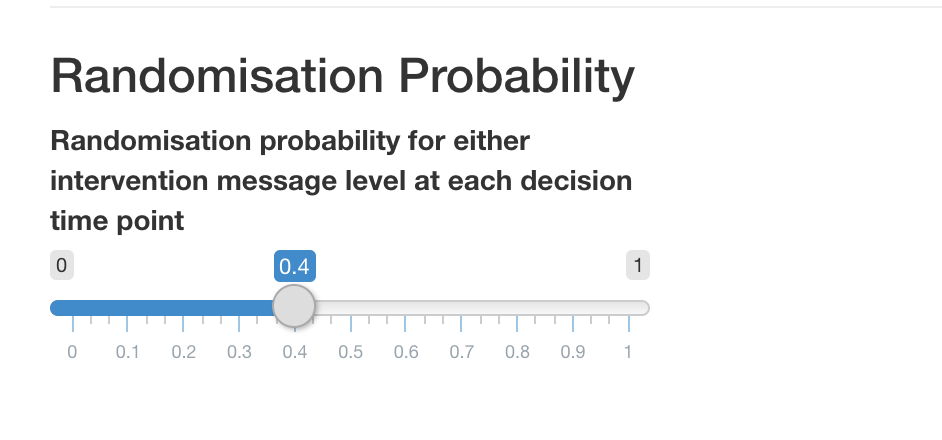} & \includegraphics[height=5cm, width=8cm]{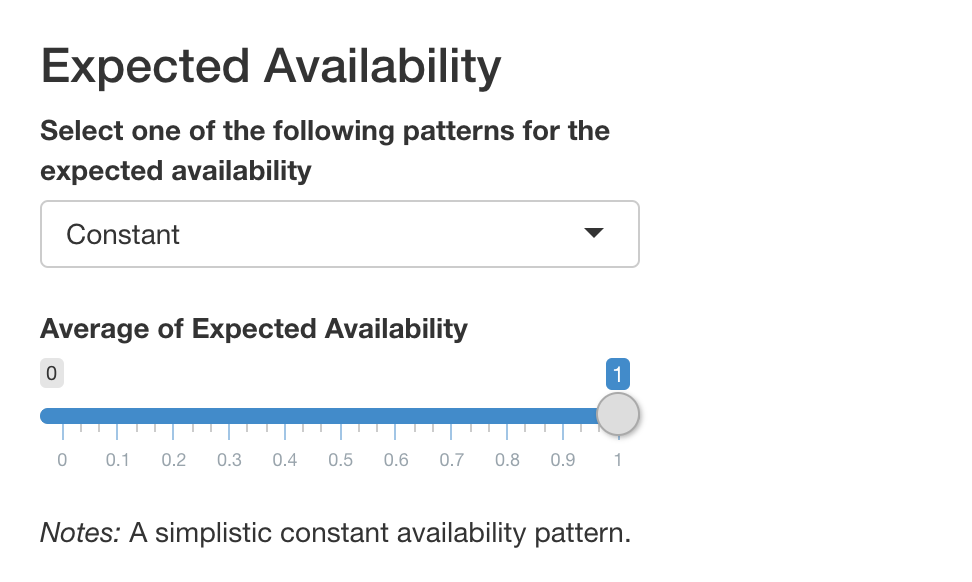} \\
 c) Randomisation probability & d) Expected availability \\
  \includegraphics[height=3cm, width=8cm]{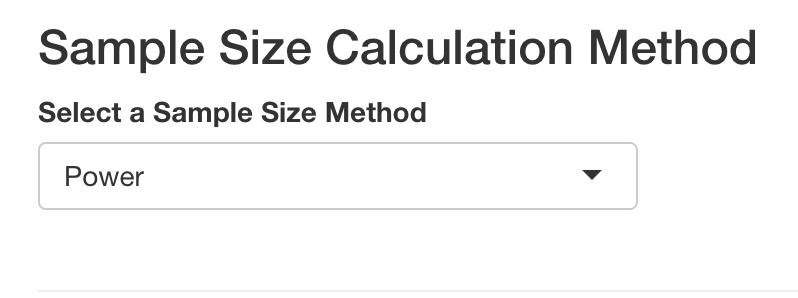} & \includegraphics[height=3cm, width=8cm]{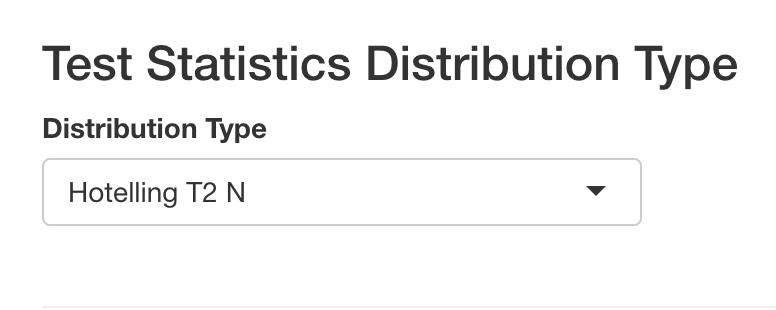} \\
 e) Sample size calculation method & f) Test statistics distribution type \\
  \includegraphics[height=6cm, width=6cm]{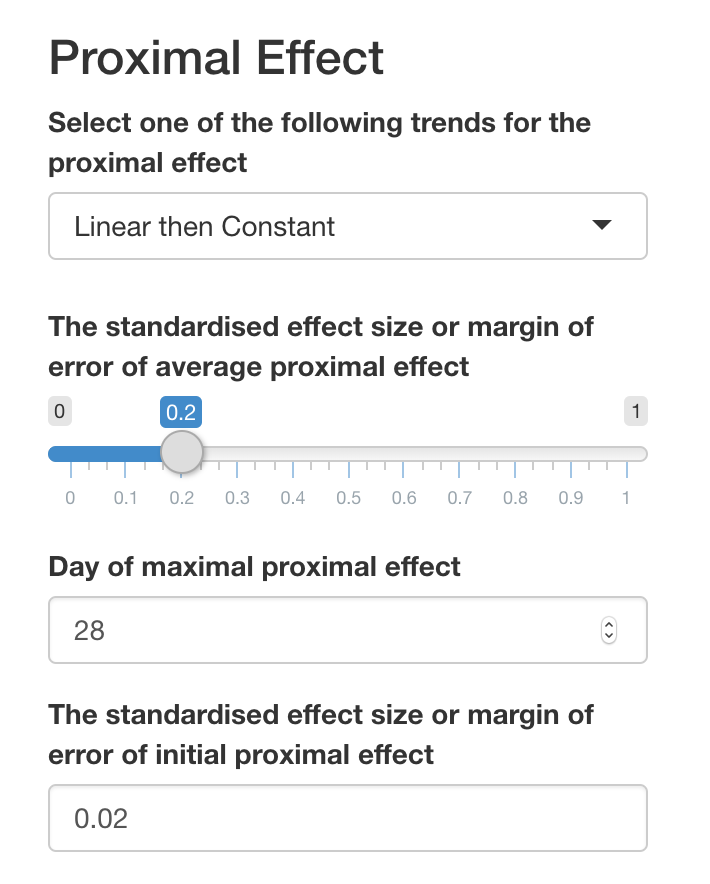} &  \includegraphics[height=6cm, width=6cm]{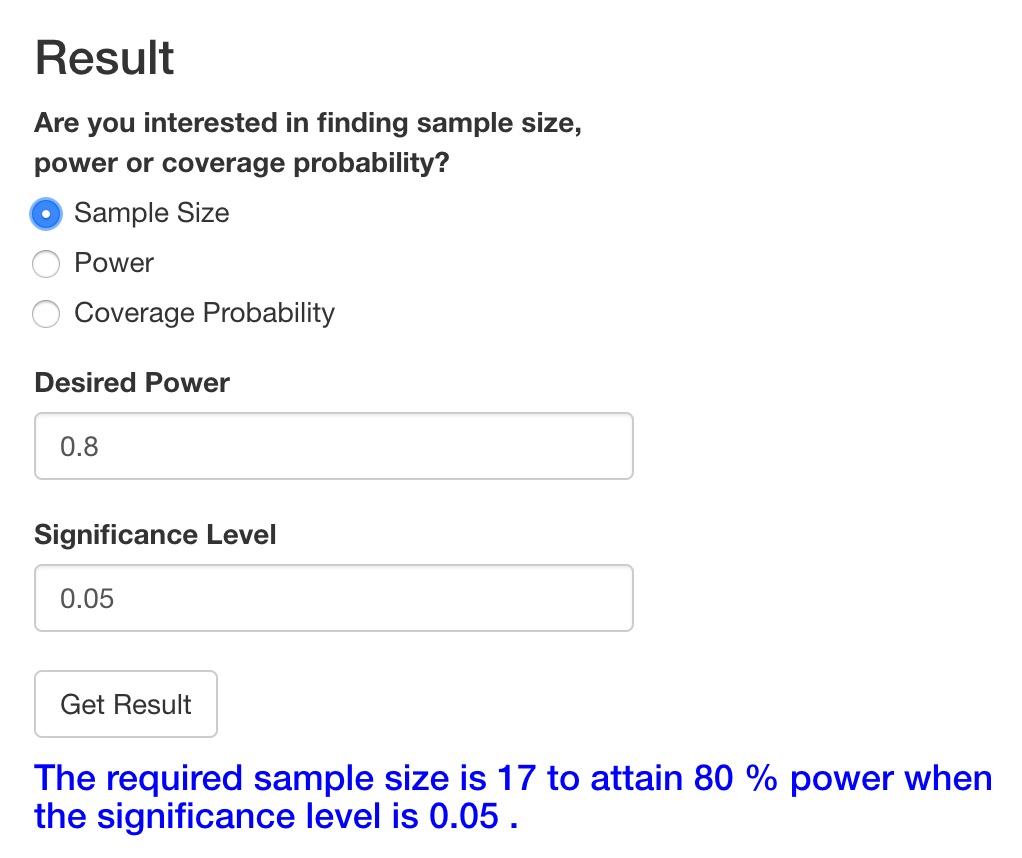} \\
 g) Proximal effect & h) Result\ 
\end{tabular}
\end{center}
\caption{
a) The study duration is $180$ days with $1$ decision time point a day. b) The number of intervention level added at the first day is $2$ and added half-way through the study in days is $2$. c) The randomisation probability for either intervention message level at each decision time point is $0.4$. d) The expected availability of each participant at each decision time point is $100\%$. e) The sample size calculation method is based on power. f) The test statistics is Hotelling's $T^2$ distributed with denominator degrees of freedom $N$. g) The proximal effect is increased linearly until 28-day then maintaining constantly at its maximum value. The initial and average values of the standardised proximal effect size for any of the intervention levels are $0.02$ and $0.2$ respectively.  h) The required sample size is calculated at a nominal power of $80\%$ and level of significance $5\%$. 
}\label{fig: shiny}
\end{figure}

\section{A Pilot Study Example}\label{s:realexample}
In this section, we demonstrate the proposed MLMRT sample size calculator for DIAMANTE study through a simple data analysis task. We use a dataset from a pilot study of the DIAMANTE project. The cohort includes 84 undergraduate and postgraduate students from the University of California, Berkeley. 
Only 22 of the sample received adaptive intervention messages while the rest 62 participants received uniform random intervention messages. 
The study period is 45 days. 
In this pilot study, the primary outcome is the change in 
daily step count between the intervention message sent day and the previous day. 
Note that the proposed sample size calculator is only applicable to the uniform random intervention group, and hence only applied there.

Here we give an example using the Motivational Message component. It has three intervention levels ($M=3$), which are all proposed before the trial. The randomization probability ($\boldsymbol\pi$) for each level (including the control level) is $0.25$.
For each of the participants, we delete the days, where the messages were not sent due to systematic errors and the outcome measures are missing. According to the working model of $Y_{id}$ defined in Section \ref{s:analysis}, we applied the constant trend of proximal effect (i.e. $q=p=1$) for the intervention levels. 
We perform the sample size calculation based on constant trend because it uses the simplest form to model the outcome measure, and gives a more intuitive estimated proximal effect sizes of the intervention levels.
The regression coefficients (i.e. $\boldsymbol\beta$) can be estimated by equation (\ref{theta_hat}). The estimates of average proximal effect sizes for the intervention levels 1-3 are $201$, $492$ and $313$ respectively. The expected conditional standard deviation of the outcome variable $\bar{\sigma}$ is $4723$. Thus, the average standardized proximal effect sizes ($\boldsymbol\beta/\bar{\sigma}$) for the intervention levels are $0.043$, $0.104$ and $0.067$ respectively.

The proximal effect sizes estimated from the pilot study can be used as input references to calculate the sample sizes required for the Motivational Message component in the uniform random intervention group in the full-fledge DIAMANTE study. 
Based on the randomization probabilities, trend and values of the standardized proximal effect sizes defined above, $100\%$ availability, $180$ days study period, Chi-square distributed test statistic, 5\% type-one error rate and 80\% power requires sample size $43$. In the similar way, based on Hotelling T-square distributed test statistic requires sample size $47$. 
In other words, for the uniform random intervention group, a message component with three intervention levels, a sample of $50$ participants is sufficient to detect the average standardized proximal effect sizes, i.e. $0.043$, $0.104$ and $0.067$, for the intervention levels 1-3 respectively, given a study period of $180$ days, at 5\% type-one error rate and 80\% power. 
Note that the DIAMANTE study aims to recruit a total of 276 patients with randomization ratio of 1:1:1. Thus there are $92$ patients allocated to the uniform random intervention group. This is higher than the sample size ($50$) estimated by the proposed sample size calculator for detecting the proximal effect sizes. 

Rather than constant trend for intervention levels of Motivational Message component, we also estimated the regression coefficients for the linear and quadratic trends assumptions.
We do not observe any statistical significant proximal effect of the intervention levels. 
However, we observe that the estimated average proximal effect of each of the intervention levels of the Motivational Message are positive. 
The linear trend gives the proximal effect estimates that show the negative associations over time. The quadratic trend estimates the proximal effect sizes that shows the concave up associations over time.
Similar analyses were also conducted for Feedback Message component, 
however the proximal effect size of the intervention levels are negative and not significant. 

\newpage
\section{Discussion}\label{s:disc}
In this paper, we propose a multi-level micro-randomised trial (MLMRT) design. 
It sequentially randomises each participant to one of the levels for each intervention message component, at each selected decision time point, over the study period. Therefore the proximal effect of each intervention level can be estimated. It is possible to allow intervention levels of some message component to be added later during the study period as a platform clinical trial. We derive the novel sample size calculation methods for MLMRT design based on not only power, but also the precision. 
For the power-based method, the required sample size is calculated in order to detect a standardized effect size at a nominal power and a significance level. For the precision-based method, the required sample size is calculated in order to achieve a precision at a nominal coverage probability when the prior knowledge of the effect size  is not given. 
Our simulation study shows that the sample sizes obtained by the proposed method give the Monte Carlo estimates of power and coverage probability close to the corresponding formulated estimates. 

Our proposed method is motivated by the Diabetes and Mental Health Adaptive Notification Tracking and Evaluation (DIAMANTE) study. 
This study is a three-arm randomized controlled trial, which include an uniform random intervention (URI) group and an adaptive intervention (AI) group. 
The MLMRT design has been applied to both the URI and AI groups. 
A physical activity application DIAMANTE has been developed. This app delivers physical activity messages, which can be improved through reinforcement learning algorithm (RLA) as in the HeartStep2 study \cite{Liao_etal_2020}.  
The participants of URI group receive message levels with equal probability while the participants of AI group receive message levels through RLA.
In this paper, the proposed sample size calculation method is only applied to the URI group. 
The calculated sample size is not aiming to select the optimal level, but it is aiming to detect whether at least one of the active intervention levels is more effective than the control level.  
However, at the stage of data analysis, we can compare the proximal main (or moderated) effects of the intervention messages at different time window over the study period.

We are going to continuously collect data through the DIAMANTE study, and perform data analysis, to further improve the experimental design. 
Though the proposed design allows new effect intervention message levels to be added half-way through the study, our design can be extended using a `decision-theoretic framework' approach similar to Lee et al (2019) \cite{Lee_etal_2019} to investigate when to add or not add a message level based on the observed proximal outcomes.
When the dataset collection is completed, we can apply linear mixed model approach to investigate the between-person heterogeneity of the proximal effect of an intervention component. This is similar to the method of Qian \textit{et al.} \cite{qian_etal_2019}. 
Based on the data analysis results of Bidargaddi \textit{et al.} \cite{Bidargaddi_etal_2018}, we could investigate that whether an intervention level of a message component is more effective at mid-day on weekends than other decision time points. The approach by Boruvka \textit{et al.} \cite{Boruvka_etal_2018} can also be used to identify the moderator effect of intervention levels of message components between the current and past decision time points on a subsequent response. 
Similar to \cite{Dempsey_etal_2020}, we can extend the proposed design to a stratified multi-level micro-randomized trial (S-ML-MRT) design.
The proposed sample size calculator of the MLMRT design can be extended to the AI group.
At each decision time point, for a particular message component, a participant is randomized to one of the message levels (both control and active interventions), determined by the proximal outcomes and message levels at previous decision time points. 
In other words, the sequential outcomes and randomization probabilities of the current message levels depend on the outcomes and message levels from previous decision time points. 
We can also name this type of design as ``multi-level micro-adaptive randomized trial''. 
In this type of design, we can estimate not only the current proximal effect, but also the delayed proximal effect, of each intervention level of a particular message component. 
Another possible future research direction is extending the sample size calculators that also involve clustered effects. For example, it is more likely that the participants' physical activity performance can be positively influenced by their group members.

\newpage
\section*{Acknowledgements}
The authors would like to thank for the valuable comments and feedback from the reviewers and editors. We also acknowledge the feedback from our colleague Dr. Raju Matti. This work has been partially supported by a grant from the Minister of Education, Singapore to Dr. Bibhas Chakraborty. The DIAMANTE trial has been funded by an R01 grant to Dr. Adrian Aguilera (University of California, Berkeley) and Dr. Courtney Lyles (University of California, San Francisco), 1R01 HS25429-01 from the Agency for Healthcare Research and Quality.


\begin{singlespace}
\bibliographystyle{acm}
\bibliography{MLMRT_KX_XY_BC}
\end{singlespace}

\section*{Appendix} \label{s:appendix}
\addcontentsline{toc}{section}{Appendix}
\renewcommand{\thesubsection}{\Alph{subsection}}

\numberwithin{equation}{section}

\subsection{\textbf{Derivation of the covariance matrix of} $\hat{\boldsymbol\beta}$}\label{app: Sigma_beta}

The second expectation of equation (\ref{theta_tilde_aa}) is
\begin{align*}
&E\left( I_{id} Y_{i, d} \boldsymbol X_{id} \right)\\
=&\begin{bmatrix} 
&E( I_{ i d } Y_{ i, d } \boldsymbol B_{d} ) \\
&E( I_{ i d } Y_{ i, d }(A_{i1d}-\pi_{1d})\boldsymbol Z_{d}\\
&\vdots\\
&E( I_{ i d }Y_{ i, d }( A_{ i M_{ 0 } d } - \pi_{ M_{ 0 } d } ) \boldsymbol Z_{d} \\
&\vdots\\
&E( I_{ i d }Y_{ i, d }( A_{ i ( \sum_{j=0}^{k-1} M_{ j } + 1 ) d } - \pi_{ ( \sum_{j=0}^{k-1} M_{ j } + 1 ) d } ) \boldsymbol Z_{d} \\
&\vdots\\
&E( I_{ i d }Y_{ i, d }( A_{ i ( \sum_{j=0}^{k} M_{ j }  ) d } - \pi_{ ( \sum_{j=0}^{k} M_{ j }  ) d } ) \boldsymbol Z_{d} \\
\end{bmatrix}\\
=&\begin{bmatrix} 
& \tau_{d} \boldsymbol B_{d}^{\top}\boldsymbol\alpha\boldsymbol B_{d} \\
& \tau_{d} E( Y_{ i, d }(A_{i1d}-\pi_{1d})\boldsymbol Z_{d}  \\
& \vdots \\
& \tau_{d} E( Y_{ i, d }( A_{ i ( \sum_{j=0}^{k} M_{ j }  ) d } - \pi_{ ( \sum_{j=0}^{k} M_{ j }  ) d } ) \boldsymbol Z_{d}\\
\end{bmatrix}
\end{align*}
where 
\begin{align*}
&E( Y_{i, d} ( A_{i1d} - \pi_{1d} ) ) \\
=& E\{ \boldsymbol B_{d}^{\top}\boldsymbol\alpha ( A_{i1d} - \pi_{1d} ) \\
+& \left[ (A_{i1d}-\pi_{1d})\boldsymbol Z_{d}^{\top}\boldsymbol\beta_1+ \cdots +(A_{iM_{0}d}-\pi_{M_{0}d})\boldsymbol Z_{d}^{\top}\boldsymbol\beta_{M_0} \right] ( A_{i1d} - \pi_{1d} ) \\
+& \left[ (A_{i( M_0 + 1 )d}-\pi_{( M_0 + 1 )d})\boldsymbol Z_{d}^{\top}\boldsymbol\beta_{ ( M_0 + 1 ) }+ \cdots +(A_{i( M_{0} + M_{1} ) d}-\pi_{( M_{0} + M_{1} )d})\boldsymbol Z_{d}^{\top}\boldsymbol\beta_{( M_{0} + M_{1} )} \right]( A_{i1d} - \pi_{1d} ) \\
&\vdots\\
+&\left[ (A_{i( \sum_{j=0}^{k-1} M_j +1) d}-\pi_{( \sum_{j=0}^{k-1}M_j + 1 )d})\boldsymbol Z_{d}^{\top}\boldsymbol\beta_{ ( \sum_{j=0}^{k-1}M_j + 1 ) }+ \cdots +(A_{i( \sum_{j=0}^{k}M_j ) d}-\pi_{( \sum_{j=0}^{k}M_j )d})\boldsymbol Z_{d}^{\top}\boldsymbol\beta_{( \sum_{j=0}^{k}M_j )} \right]\\ 
&( A_{i1d} - \pi_{1d} ) \\
+&\epsilon_{id} ( A_{i1d} - \pi_{1d} ) \} \\
=& \left[ \pi_{ 1 d }(1 - \pi_{ 1 d } )\boldsymbol Z_{d}^{\top}\boldsymbol\beta_1- \cdots - \pi_{ 1 d }\pi_{ M_{0} d }\boldsymbol Z_{d}^{\top}\boldsymbol\beta_{M_0} \right]\\
+& \left[  - \pi_{ 1 d }\pi_{ ( M_0 + 1 ) d } \boldsymbol Z_{d}^{\top}\boldsymbol\beta_{ ( M_0 + 1 ) } - \cdots - \pi_{ 1 d }\pi_{ ( M_{0} + M_{1} ) d }\boldsymbol Z_{d}^{\top}\boldsymbol\beta_{( M_{0} + M_{1} )} \right] \\
&\vdots\\
+& \left[ - \pi_{ 1 d }\pi_{ ( \sum_{j=0}^{k-1}M_j + 1 ) d }\boldsymbol Z_{d}^{\top}\boldsymbol\beta_{ ( \sum_{j=0}^{k-1}M_j + 1 ) } - \cdots - \pi_{ 1 d }\pi_{ ( \sum_{j=0}^{k}M_j )d }\boldsymbol Z_{d}^{\top}\boldsymbol\beta_{ ( \sum_{j=0}^{k}M_j ) } \right], \\
\end{align*} 
hence, in the similar way, 
\begin{align*}
& E( Y_{i, d}( A_{ i ( \sum_{j=0}^{k} M_{ j }  ) d } - \pi_{ ( \sum_{j=0}^{k} M_{ j }  ) d } ) ) \\
=  & \left[ -\pi_{ ( \sum_{j=0}^{k}M_j ) d }\pi_{ 1 d }\boldsymbol Z_{d}^{\top}\boldsymbol\beta_1- \cdots - \pi_{ ( \sum_{j=0}^{k}M_j ) d }\pi_{ M_{0} d }\boldsymbol Z_{d}^{\top}\boldsymbol\beta_{M_0} \right] \\
+& \left[  - \pi_{ ( \sum_{j=0}^{k}M_j ) d }\pi_{ ( M_0 + 1 ) d } \boldsymbol Z_{d}^{\top}\boldsymbol\beta_{ ( M_0 + 1 ) } - \cdots - \pi_{ ( \sum_{j=0}^{k}M_j ) d }\pi_{ ( M_{0} + M_{1} ) d }\boldsymbol Z_{d}^{\top}\boldsymbol\beta_{( M_{0} + M_{1} )} \right] \\
&\vdots\\
+& \left[ -\pi_{ ( \sum_{j=0}^{k}M_j ) d }\pi_{ ( \sum_{j=0}^{k-1}M_j + 1 ) d } \boldsymbol Z_{d}^{\top}\boldsymbol\beta_{ ( \sum_{j=0}^{k-1}M_j + 1 ) } - \cdots + \pi_{ ( \sum_{j=0}^{k}M_j ) d }( 1- \pi_{ ( \sum_{j=0}^{k}M_j )d } )\boldsymbol Z_{d}^{\top}\boldsymbol\beta_{ ( \sum_{j=0}^{k}M_j ) } \right]. \\
\end{align*}

The first expectation of equation (\ref{theta_tilde_aa}) is
\begin{align*}
& E\left( I_{id} \boldsymbol X_{id} \boldsymbol X_{id}^{\top} \right)\\
=& \tau_{ d } E\lbrace \begin{bmatrix}  
& \boldsymbol B_{d} \\
&  ( A_{ i 1 d } - \pi_{ 1 d } ) \boldsymbol Z_{d}  \\
& \vdots  \\
& ( A_{ i \sum_{ j=0 }^{ k } M_{ j } d } - \pi_{ \sum_{ j=0 }^{ k } M_{ j } d } ) \boldsymbol Z_{d}  \\
\end{bmatrix} 
\begin{bmatrix}  
& \boldsymbol B_{d}^{\top}, ( A_{ i 1 d } - \pi_{ 1 d } ) \boldsymbol Z_{d}^{\top} , \cdots, ( A_{ i \sum_{j=0}^{k}M_{j} d } - \pi_{ \sum_{j=0}^{k} M_{j} d } ) \boldsymbol Z_{d}^{\top}  \\
\end{bmatrix} \rbrace \\
=& \tau_{ d } E \begin{bmatrix}
\boldsymbol B_{d} \boldsymbol B_{d}^{\top} & 
\cdots &
(A_{i \sum_{j=0}^{k}M_j d}-\pi_{\sum_{j=0}^{k}M_j d})\boldsymbol B_{d}\boldsymbol Z_{d}^{\top}
\\
\vdots & 
\ddots &
\vdots
\\
(A_{i \sum_{j=0}^{k}M_j d}-\pi_{ \sum_{j=0}^{k}M_j d})\boldsymbol Z_{d} \boldsymbol B_{d}^{\top} &
\cdots &
(A_{i \sum_{j=0}^{k}M_j d}-\pi_{ \sum_{j=0}^{k}M_j d})(A_{i \sum_{j=0}^{k}M_j d}-\pi_{ \sum_{j=0}^{k}M_j d})\boldsymbol Z_{d}\boldsymbol Z_{d}^{\top}
\end{bmatrix}
\\
=&\tau_{d} \begin{bmatrix}
\boldsymbol B_{d} \boldsymbol B_{d}^{\top} & 
0 & 
\cdots & 
0
\\
0 & 
\pi_{1d}(1-\pi_{1d})\boldsymbol Z_{d} \boldsymbol Z_{d}^{\top} &
\cdots &
- \pi_{1d}\pi_{ ( \sum_{j=0}^{k}M_j ) d}\boldsymbol Z_{d} \boldsymbol Z_{d}^{\top}
\\ 
\vdots & 
\vdots &
\ddots &
\vdots
\\
0 & 
- \pi_{ ( \sum_{j=0}^{k}M_j ) d}\pi_{1d}\boldsymbol Z_{d} \boldsymbol Z_{d}^{\top} & 
\cdots & 
 \pi_{ ( \sum_{j=0}^{k}M_j ) d}(1-\pi_{ ( \sum_{j=0}^{k}M_j ) d})\boldsymbol Z_{d} \boldsymbol Z_{d}^{\top}
\\
\end{bmatrix}.
\end{align*}
The estimators from (\ref{theta_tilde_aa}) are
\begin{equation}
\tilde{\boldsymbol\alpha}=\left( \sum_{d=1}^D\tau_{d}\boldsymbol B_{d}\boldsymbol B_{d}^{\top} \right)^{-1} \sum_{d=1}^D\tau_{d}\boldsymbol B_{d}^{\top}\boldsymbol\alpha\boldsymbol B_{d}
\end{equation}
and

\begin{align*}
\tilde{\boldsymbol\beta}&=(\tilde{\boldsymbol\beta}_{1}^{\top},\ldots,\tilde{\boldsymbol\beta}_{M}^{\top})^{\top}\\
=\sum_{d=1}^D\tau_{d} &\begin{bmatrix}
\pi_{1d}(1-\pi_{1d})\boldsymbol Z_{d} \boldsymbol Z_{d}^{\top} &
\cdots &
 - \pi_{1d}\pi_{ ( \sum_{j=0}^{k}M_j ) d}\boldsymbol Z_{d} \boldsymbol Z_{d}^{\top}
\\ 
\vdots & 
\ddots &
\vdots
\\
 - \pi_{ ( \sum_{j=0}^{k}M_j ) d}\pi_{1d}\boldsymbol Z_{d} \boldsymbol Z_{d}^{\top} &  
\cdots & 
 \pi_{ ( \sum_{j=0}^{k}M_j ) d}(1-\pi_{ ( \sum_{j=0}^{k}M_j ) d})\boldsymbol Z_{d} \boldsymbol Z_{d}^{\top}
\\
\end{bmatrix}^{-1}\\
\sum_{d=1}^D\tau_{d} &\begin{bmatrix}
\left[ \pi_{1d}(1-\pi_{1d})\boldsymbol Z_{d}^{\top}\boldsymbol\beta_1  - \cdots - \pi_{1d}\pi_{ ( \sum_{j=0}^{k}M_j ) d}\boldsymbol Z_{d}^{\top}\boldsymbol\beta_{ ( \sum_{j=0}^{k}M_j ) }  \right] \boldsymbol Z_{d}\\
\vdots\\
 \left[ -\pi_{1d}\pi_{ ( \sum_{j=0}^{k}M_j ) d}\boldsymbol Z_{d}^{\top}\boldsymbol\beta_1  - \cdots + \pi_{ ( \sum_{j=0}^{k}M_j ) d}(1-\pi_{ ( \sum_{j=0}^{k}M_j ) d})\boldsymbol Z_{d}^{\top}\boldsymbol\beta_{ ( \sum_{j=0}^{k}M_j ) }  \right] \boldsymbol Z_{d}\\
\end{bmatrix}.
\end{align*}

\begin{align*}
&\sqrt{N}(\hat{\boldsymbol\theta}-\tilde{\boldsymbol\theta})\\
=&\sqrt{N}\left\lbrace \left[ \dfrac{1}{N}\sum_{i=1}^{N}\sum_{d=1}^{D}I_{id}\boldsymbol X_{id} \boldsymbol X_{id}^{\top} \right]^{-1}\dfrac{1}{N}\sum_{i=1}^{N}\sum_{d=1}^{D}I_{id}Y_{i, d}\boldsymbol X_{id} - \tilde{\boldsymbol\theta} \right\rbrace\\
=&\sqrt{N}\left\lbrace \left[ \dfrac{1}{N}\sum_{i=1}^{N}\sum_{d=1}^{D}I_{id}\boldsymbol X_{id} \boldsymbol X_{id}^{\top} \right]^{-1} \dfrac{1}{N}\sum_{i=1}^{N}\sum_{d=1}^{D}\left[ I_{id}Y_{i, d}\boldsymbol X_{id} - I_{id}\boldsymbol X_{id} \boldsymbol X_{id}^{\top} \tilde{\boldsymbol\theta} \right] \right\rbrace\\
=&\sqrt{N}\left\lbrace \left[ \sum_{d=1}^{D}E( I_{id}\boldsymbol X_{id} \boldsymbol X_{id}^{\top}) \right]^{-1} \dfrac{1}{N}\sum_{i=1}^{N}\sum_{d=1}^{D}\left[ I_{id}\boldsymbol X_{id} \tilde{\epsilon}_{id} \right] \right\rbrace + O_p(1),\\
\end{align*}
where $O_p(1)\xrightarrow{N\rightarrow\infty}0$, $\tilde{\epsilon}_{id}$=$Y_{i, d}$ - $\boldsymbol B_{d}^{\top}\tilde{\boldsymbol\alpha}-(A_{i1d}-\pi_{1d})\boldsymbol Z_{d}^{\top}\tilde{\boldsymbol\beta}_1- \cdots -(A_{iMd}-\pi_{Md})\boldsymbol Z_{d}^{\top}\tilde{\boldsymbol\beta}_M$ and $E( I_{id}\boldsymbol X_{id} \tilde{\epsilon}_{id})$=0.

\begin{align*}
&E(\sum_{d=1}^{D} I_{id}\tilde{\epsilon}_{id}\boldsymbol X_{id}  \sum_{d=1}^{D} I_{id}\tilde{\epsilon}_{id}\boldsymbol X_{id}^{\top} )\\
=&E(\sum_{d=1}^{D} I_{id}\tilde{\epsilon}_{id}^2\boldsymbol X_{id}\boldsymbol X_{id}^{\top} + \sum_{(d)\neq(d)^\prime}^{(D)} I_{i(d)}I_{i(d)^\prime}\tilde{\epsilon}_{i(d)} \tilde{\epsilon}_{i(d)^\prime} \boldsymbol X_{i(d)}\boldsymbol X_{i(d)^\prime}^{\top})\\
=&\bar{\sigma}^2\sum_{d=1}^{D} \tau_{d}\begin{bmatrix}
 \boldsymbol B_{d} \boldsymbol B_{d}^{\top} & 
0 & 
\cdots & 
0
\\
0 & 
 \pi_{1d}(1-\pi_{1d})\boldsymbol Z_{d} \boldsymbol Z_{d}^{\top} &
\cdots &
- \pi_{1d}\pi_{ ( \sum_{j=0}^{k}M_j  ) d}\boldsymbol Z_{d} \boldsymbol Z_{d}^{\top} 
\\ 
\vdots & 
\vdots &
\ddots &
\vdots
\\
0 & 
- \pi_{ ( \sum_{j=0}^{k}M_j  ) dt}\pi_{1d}\boldsymbol Z_{d} \boldsymbol Z_{d}^{\top}  & 
\cdots & 
 \pi_{ ( \sum_{j=0}^{k}M_j ) d}(1-\pi_{ ( \sum_{j=0}^{k}M_j  ) d})\boldsymbol Z_{d}\boldsymbol Z_{d}^{\top} 
\\
\end{bmatrix}\\
+& \sum_{(d)\neq (d)^\prime}^{(D)} \sigma_{(d)(d)^\prime}\tau_{d}\tau_{(d)^\prime} \begin{bmatrix}
\boldsymbol B_{d} \boldsymbol B_{(d)^\prime}^{\top} & 
0 & 
\cdots & 
0
\\
0 & 
0 &
\cdots &
0
\\ 
\vdots & 
\vdots &
\ddots &
\vdots
\\
0 & 
0 & 
\cdots & 
0
\\
\end{bmatrix}.
\end{align*}

The asymptotic covariance matrix of $\hat{\boldsymbol\beta}$ can be defined by
\begin{align*}
\boldsymbol\Sigma_{\boldsymbol\beta\boldsymbol\beta}=&\sum_{d=1}^D\tau_{d}\begin{bmatrix}
 \pi_{1d}(1-\pi_{1d})\boldsymbol Z_{d} \boldsymbol Z_{d}^{\top} &
\cdots &
 -\pi_{1d}\pi_{ ( \sum_{j=0}^{k}M_j ) d}\boldsymbol Z_{d} \boldsymbol Z_{d}^{\top} 
\\ 
\vdots & 
\ddots &
\vdots
\\
 - \pi_{ ( \sum_{j=0}^{k}M_j ) d}\pi_{1d}\boldsymbol Z_{d} \boldsymbol Z_{d}^{\top} &  
\cdots & 
 \pi_{ ( \sum_{j=0}^{k}M_j ) d}(1-\pi_{ ( \sum_{j=0}^{k}M_j ) d})\boldsymbol Z_{d} \boldsymbol Z_{d}^{\top}
\\
\end{bmatrix}^{-1}\\
\bar{\sigma}^2&\sum_{d=1}^D\tau_{d}\begin{bmatrix}
 \pi_{1d}(1-\pi_{1d})\boldsymbol Z_{d} \boldsymbol Z_{d}^{\top} &
\cdots &
 - \pi_{1d}\pi_{ ( \sum_{j=0}^{k}M_j ) d}\boldsymbol Z_{d} \boldsymbol Z_{d}^{\top}
\\ 
\vdots & 
\ddots &
\vdots
\\
 - \pi_{ ( \sum_{j=0}^{k}M_j ) d}\pi_{1d}\boldsymbol Z_{d} \boldsymbol Z_{d}^{\top} &  
\cdots & 
 \pi_{ ( \sum_{j=0}^{k}M_j ) d}(1-\pi_{ ( \sum_{j=0}^{k}M_j ) d})\boldsymbol Z_{d} \boldsymbol Z_{d}^{\top}
\\
\end{bmatrix}\\
&\sum_{d=1}^D\tau_{d}\begin{bmatrix}
 \pi_{1d}(1-\pi_{1d})\boldsymbol Z_{d} \boldsymbol Z_{d}^{\top} &
\cdots &
 - \pi_{1d}\pi_{ ( \sum_{j=0}^{k}M_j ) d}\boldsymbol Z_{d} \boldsymbol Z_{d}^{\top}
\\ 
\vdots & 
\ddots &
\vdots
\\
 - \pi_{ ( \sum_{j=0}^{k}M_j ) d}\pi_{1d}\boldsymbol Z_{d} \boldsymbol Z_{d}^{\top} &  
\cdots & 
 \pi_{ ( \sum_{j=0}^{k}M_j ) d}(1-\pi_{ ( \sum_{j=0}^{k}M_j ) d})\boldsymbol Z_{d} \boldsymbol Z_{d}^{\top}
\\
\end{bmatrix}^{-1}\\
=\bar{\sigma}^2&\sum_{d=1}^D\tau_{d}\begin{bmatrix}
 \pi_{1d}(1-\pi_{1d})\boldsymbol Z_{d} \boldsymbol Z_{d}^{\top} &
\cdots &
 - \pi_{1d}\pi_{ ( \sum_{j=0}^{k}M_j ) d}\boldsymbol Z_{d} \boldsymbol Z_{d}^{\top}
\\ 
\vdots & 
\ddots &
\vdots
\\
 - \pi_{ ( \sum_{j=0}^{k}M_j ) d}\pi_{1d}\boldsymbol Z_{d} \boldsymbol Z_{d}^{\top} &  
\cdots & 
 \pi_{ ( \sum_{j=0}^{k}M_j ) d}(1-\pi_{ ( \sum_{j=0}^{k}M_j ) d})\boldsymbol Z_{d} \boldsymbol Z_{d}^{\top}
\\
\end{bmatrix}^{-1}, 
\end{align*}
where $\bar{\sigma}^2$=$\sum_{d=1}^{D}E(\text{Var}(Y_{i, d}\mid I_{id}=1, A_{i1d},\ldots,A_{i \sum_{j=0}^{k}M_j d} ))/(D)$.

\subsection{\textbf{Proof of Theorem \ref{theo1}}}\label{app: ThePro}

\noindent \textbf{Proof:} The proof of consistency requires the result of strong law of large numbers, such that 
 $\hat{\boldsymbol\theta}\rightarrow\tilde{\boldsymbol\theta}$, almost surely, and uniformly for $\boldsymbol\theta\in\boldsymbol\Theta$} as $N\rightarrow\infty$ and $\tilde{\boldsymbol\theta}$ being the unique zero value of $E(LSE(\boldsymbol\theta))$ due to assumption g). To prove the asymptotic normality result, by the central limit theorem, $\sqrt{N}(\hat{\boldsymbol\theta}-\tilde{\boldsymbol\theta})$ converges in distribution to $\text{Normal}(0,\boldsymbol\Sigma_{\boldsymbol\theta})$. Note that $\boldsymbol\Sigma_{\boldsymbol\theta}$ is derived in Appendix \ref{app: Sigma_beta}.
\\

\subsection{\textbf{Derivation of the test statistic distribution under a small sample}}\label{app: DistrTest}

For the small $N$ estimator of $\hat{\boldsymbol\Sigma}_{\boldsymbol\beta\boldsymbol\beta}$, we assume the following two approximations based on \cite{Mancl_DeRouen_2001}, i.e.
\begin{align*}
E(\boldsymbol \hat{e}_{i} \boldsymbol \hat{e}_{i}^{\top}) \approx (\boldsymbol I_{D\times D} - \boldsymbol H_i)\text{COV}(\boldsymbol Y_i)(\boldsymbol I_{D\times D} - \boldsymbol H_i)^{\top} 
\end{align*}
and
\begin{align*}
&\sum_{i=1}^N \boldsymbol X_{i} (\boldsymbol I_{D\times D} - \boldsymbol H_i)^{-1} \hat{\boldsymbol e}_{i}\hat{\boldsymbol e}_{i}^{\top} (\boldsymbol I_{D\times D} - \boldsymbol H_i)^{-1} \boldsymbol X_{i}^{\top}\\
\approx & \sum_{i=1}^N \boldsymbol X_{i} \text{COV}(\boldsymbol Y_i) \boldsymbol X_{i}^{\top}\\
\approx & \sum_{i=1}^N \boldsymbol X_{i} \boldsymbol \epsilon_{i} \boldsymbol \epsilon_{i}^{\top} \boldsymbol X_{i}^{\top},
\end{align*}
where
\begin{align*}
\boldsymbol X_{i} \boldsymbol \epsilon_{i}  =
\begin{bmatrix}
\sum_{d=1}^{D}I_{id}  \epsilon_{id} \\
\vdots\\
\sum_{d=1}^{D} I_{id} ( d - 1 )^{q-1} \epsilon_{id} \\
\sum_{d=1}^{D} I_{id} ( A_{i1d} - \pi_{1d} ) \epsilon_{id} \\
\vdots\\
\sum_{d=1}^{D} I_{id} ( A_{i1d} - \pi_{1d} ) ( d - 1 )^{p-1} \epsilon_{id} \\
\vdots\\
\sum_{d=1}^{D} I_{id} ( A_{iM_{0}d} - \pi_{M_{0}d} ) \epsilon_{id} \\
\vdots\\
\sum_{d=1}^{D} I_{id} ( A_{iM_{0}d} - \pi_{M_{0}d} ) ( d - 1 )^{p-1} \epsilon_{id} \\
\vdots\\
\sum_{d=1}^{D} I_{id} ( A_{i\sum_{j=0}^{k}M_{j}d} - \pi_{\sum_{j=0}^{k}M_{j}d} ) \epsilon_{id} \\
\vdots\\
\sum_{d=1}^{D} I_{id} ( A_{i\sum_{j=0}^{k}M_{j}d} - \pi_{\sum_{j=0}^{k}M_{j}d} ) ( d - 1 )^{p-1} \epsilon_{id} \\
\end{bmatrix}
\end{align*}
with dimension $\left( q + p\sum_{j=0}^k M_j \right) \times 1$. Hence, we can have $\hat{\boldsymbol\Sigma}_{\boldsymbol\beta\boldsymbol\beta}$ approximated by 
\begin{align*}
\boldsymbol U
\left( \sum_{i=1}^N \boldsymbol X_{i} \boldsymbol X_{i}^{\top}/N \right)^{-1}
\left( \sum_{i=1}^N \boldsymbol X_{i} \boldsymbol \epsilon_{i} \boldsymbol \epsilon_{i}^{\top} \boldsymbol X_{i}^{\top} \right)
\left( \sum_{i=1}^N \boldsymbol X_{i} \boldsymbol X_{i}^{\top}/N \right)^{-1}
\boldsymbol U^{\top},
\end{align*}
where $\boldsymbol U$ is a rectangular matrix with dimension $\left( p\sum_{j=0}^k M_j \right) \times \left( q + p\sum_{j=0}^k M_j \right)$,  i.e. $U = \left[ \boldsymbol U_0, \boldsymbol U_1 \right]$, where $\boldsymbol U_0$ is a zeros matrix with dimension $p\sum_{j=0}^k M_j \times q$ while $\boldsymbol U_1$ is an identity matrix with dimension $p\sum_{j=0}^k M_j \times p\sum_{j=0}^k M_j$ .
Assuming $\boldsymbol X_{i}$ are given, but not $\boldsymbol\epsilon_i$ and hence $\boldsymbol X_{i} \boldsymbol \epsilon_{i}$ are random, for $i=1,\ldots N$. Thus, the degrees of freedom for $\boldsymbol X_{i} \boldsymbol \epsilon_{i}$ can be $N$ when assuming no restrictions on $\boldsymbol X_{i} \boldsymbol \epsilon_{i}$. Alternatively, the degrees of freedom for $\boldsymbol X_{i} \boldsymbol \epsilon_{i}$ can be $N-1$ when assuming a restriction, e.g.,  $\sum_{i=1}^N\boldsymbol X_{i} \boldsymbol \epsilon_{i}=\boldsymbol 0$.

\subsection{\textbf{Further simulation results}}\label{app: FutherSim}

\renewcommand{\thetable}{C.\arabic{table}}

\newpage
\begin{table}[H]
\caption{Sample sizes calculation based on power (P) when the standardized proximal effect size of intervention levels satisfy $\delta_m(d)=\boldsymbol Z_{d}^{\top}\boldsymbol \delta_m$, where $\boldsymbol \delta_m=\boldsymbol\beta_m/\sigma$, for $m=1,\ldots,M$. Note that we have $M=3$, 4, $\sigma=1$ and $\rho=0$. The significance level is 0.05. The desired P is 0.80. ``Duration" is the duration of study ($D$) in days. 
Constant trend for proximal effect size and $100\%$ availability at each time point are assumed.
The initial standardized proximal effect size is 0.02.
}
\label{Table: TC5}
\centering
\begin{tabular}{rllrrrrrrr}
  \hline
 & \multicolumn{3}{c}{} &  \multicolumn{2}{c}{ Sample Size } & \multicolumn{2}{c}{ Formulated P }  & \multicolumn{2}{c}{ Monte Carlo P } \\
 & \multicolumn{3}{c}{} &  \multicolumn{6}{c}{ Average standardized proximal effect size}  \\
& Test Statistics & M & Duration & 0.20 & 0.10 & 0.20 & 0.10 & 0.20 & 0.10 \\ 
  \hline
& $\chi^2_{Mp}$ & 3 & 180 & 7.00 & 26.00 & 0.84 & 0.81 & 0.83 & 0.81 \\ 
 &  &   &  84 & 14.00 & 55.00 & 0.82 & 0.81 & 0.79 & 0.80 \\ 
 &  &   &  28 & 41.00 & 163.00 & 0.80 & 0.80 & 0.81 & 0.80 \\ 
 &  &   &  14 & 82.00 & 325.00 & 0.80 & 0.80 & 0.79 & 0.79 \\ 
 & Hotelling's $T^2_{Mp, N}$ &   & 180 & 11.00 & 30.00 & 0.85 & 0.81 & 0.84 & 0.79 \\ 
 &  &   &  84 & 18.00 & 58.00 & 0.82 & 0.80 & 0.80 & 0.78 \\ 
 &  &   &  28 & 45.00 & 167.00 & 0.81 & 0.80 & 0.82 & 0.82 \\ 
 &  &   &  14 & 86.00 & 329.00 & 0.81 & 0.80 & 0.79 & 0.79 \\ 
 & Hotelling's $T^2_{Mp, N-1}$ &   & 180 & 11.00 & 30.00 & 0.82 & 0.81 & 0.78 & 0.78 \\ 
 &  &   &  84 & 18.00 & 59.00 & 0.81 & 0.81 & 0.77 & 0.78 \\ 
 &  &   &  28 & 45.00 & 167.00 & 0.80 & 0.80 & 0.78 & 0.80 \\ 
 &  &   &  14 & 86.00 & 329.00 & 0.80 & 0.80 & 0.79 & 0.79 \\ 
 & Hotelling's $T^2_{Mp, N-q-1}$ &   & 180 & 12.00 & 30.00 & 0.86 & 0.81 & 0.82 & 0.79 \\ 
 &  &   &  84 & 18.00 & 59.00 & 0.80 & 0.81 & 0.76 & 0.81 \\ 
 &  &   &  28 & 45.00 & 167.00 & 0.80 & 0.80 & 0.79 & 0.82 \\ 
 &  &   &  14 & 86.00 & 329.00 & 0.80 & 0.80 & 0.79 & 0.79 \\ 
 & $\chi^2_{Mp}$ & 4 & 180 & 7.00 & 28.00 & 0.81 & 0.81 & 0.80 & 0.80 \\ 
 &  &   &  84 & 15.00 & 60.00 & 0.81 & 0.81 & 0.80 & 0.80 \\ 
 &  &   &  28 & 45.00 & 178.00 & 0.81 & 0.80 & 0.80 & 0.81 \\ 
 &  &   &  14 & 89.00 & 356.00 & 0.80 & 0.80 & 0.79 & 0.82 \\ 
 & Hotelling's $T^2_{Mp, N}$ &   & 180 & 12.00 & 33.00 & 0.81 & 0.81 & 0.80 & 0.80 \\ 
 &  &   &  84 & 20.00 & 64.00 & 0.81 & 0.80 & 0.80 & 0.81 \\ 
 &  &   &  28 & 50.00 & 183.00 & 0.81 & 0.80 & 0.80 & 0.81 \\ 
 &  &   &  14 & 94.00 & 360.00 & 0.80 & 0.80 & 0.80 & 0.78 \\ 
 & Hotelling's $T^2_{Mp, N-1}$ &   & 180 & 13.00 & 33.00 & 0.85 & 0.81 & 0.79 & 0.78 \\ 
 &  &   &  84 & 20.00 & 65.00 & 0.80 & 0.81 & 0.76 & 0.82 \\ 
 &  &   &  28 & 50.00 & 183.00 & 0.81 & 0.80 & 0.78 & 0.81 \\ 
 &  &   &  14 & 94.00 & 360.00 & 0.80 & 0.80 & 0.80 & 0.80 \\ 
 & Hotelling's $T^2_{Mp, N-q-1}$ &   & 180 & 13.00 & 33.00 & 0.82 & 0.80 & 0.78 & 0.82 \\ 
 &  &   &  84 & 21.00 & 65.00 & 0.82 & 0.81 & 0.80 & 0.79 \\ 
 &  &   &  28 & 50.00 & 183.00 & 0.81 & 0.80 & 0.80 & 0.80 \\ 
 &  &   &  14 & 94.00 & 360.00 & 0.80 & 0.80 & 0.77 & 0.76 \\     \hline
\end{tabular}
\end{table}

\begin{table}[H]
\caption{Sample sizes calculation based on coverage probability (CP) when the standardized proximal effect size of intervention levels satisfy $\delta_m(d)=\boldsymbol Z_{d}^{\top}\boldsymbol \delta_m$, where $\boldsymbol \delta_m=\boldsymbol\beta_m/\sigma$, for $m=1,\ldots,M$. Note that we have $M=3$, 4, $\sigma=1$ and $\rho=0$. 
The desired CP is $95\%$.
 ``Duration" is the duration of study ($D$) in days. 
 Constant trend for standardized proximal effect size and $100\%$ availability at each time point are assumed.
Precision of initial standardized proximal effect size is 0.02.
}
\label{Table: TC6}
\centering
\begin{tabular}{rllrrrrrrr}
  \hline
  & \multicolumn{3}{c}{} & \multicolumn{2}{c}{ Sample Size } & \multicolumn{2}{c}{ Formulated CP }  & \multicolumn{2}{c}{ Monte Carlo CP } \\
 & \multicolumn{3}{c}{} &  \multicolumn{6}{c}{Precision of average standardized proximal effect size}  \\
 & Test Statistics & M & Duration & 0.25 & 0.15 & 0.25 & 0.15 & 0.25 & 0.15 \\ 
  \hline
& $\chi^2_{Mp}$ & 3 & 180 & 5.00 & 19.00 & 0.97 & 0.96 & 0.96 & 0.96 \\ 
 &  &   &  84 & 10.00 & 39.00 & 0.96 & 0.95 & 0.96 & 0.94 \\ 
 &  &   &  28 & 30.00 & 117.00 & 0.96 & 0.95 & 0.95 & 0.95 \\ 
 &  &   &  14 & 59.00 & 233.00 & 0.95 & 0.95 & 0.95 & 0.95 \\ 
 & Hotelling's $T^2_{Mp, N}$ &   & 180 & 10.00 & 24.00 & 0.96 & 0.96 & 0.97 & 0.96 \\ 
 &  &   &  84 & 15.00 & 45.00 & 0.95 & 0.95 & 0.96 & 0.96 \\ 
 &  &   &  28 & 35.00 & 122.00 & 0.95 & 0.95 & 0.95 & 0.95 \\ 
 &  &   &  14 & 64.00 & 238.00 & 0.95 & 0.95 & 0.94 & 0.96 \\ 
 & Hotelling's $T^2_{Mp, N-1}$ &   & 180 & 10.00 & 24.00 & 0.95 & 0.95 & 0.96 & 0.96 \\ 
 &  &   &  84 & 16.00 & 45.00 & 0.96 & 0.95 & 0.97 & 0.95 \\ 
 &  &   &  28 & 35.00 & 122.00 & 0.95 & 0.95 & 0.96 & 0.95 \\ 
 &  &   &  14 & 64.00 & 238.00 & 0.95 & 0.95 & 0.96 & 0.95 \\ 
 & Hotelling's $T^2_{Mp, N-q-1}$ &   & 180 & 11.00 & 24.00 & 0.96 & 0.95 & 0.97 & 0.95 \\ 
 &  &   &  84 & 16.00 & 45.00 & 0.96 & 0.95 & 0.96 & 0.95 \\ 
 &  &   &  28 & 35.00 & 122.00 & 0.95 & 0.95 & 0.96 & 0.95 \\ 
 &  &   &  14 & 64.00 & 239.00 & 0.95 & 0.95 & 0.95 & 0.95 \\ 
 & $\chi^2_{Mp}$  & 4 & 180 & 6.00 & 22.00 & 0.97 & 0.95 & 0.97 & 0.95 \\ 
 &  &   &  84 & 12.00 & 48.00 & 0.95 & 0.95 & 0.95 & 0.94 \\ 
 &  &   &  28 & 36.00 & 142.00 & 0.95 & 0.95 & 0.96 & 0.94 \\ 
 &  &   &  14 & 71.00 & 283.00 & 0.95 & 0.95 & 0.94 & 0.93 \\ 
 & Hotelling's $T^2_{Mp, N}$ &   & 180 & 12.00 & 29.00 & 0.96 & 0.95 & 0.97 & 0.96 \\ 
 &  &   &  84 & 19.00 & 54.00 & 0.96 & 0.95 & 0.96 & 0.95 \\ 
 &  &   &  28 & 42.00 & 148.00 & 0.95 & 0.95 & 0.96 & 0.95 \\ 
 &  &   &  14 & 78.00 & 290.00 & 0.95 & 0.95 & 0.95 & 0.96 \\ 
 & Hotelling's $T^2_{Mp, N-1}$ &   & 180 & 13.00 & 29.00 & 0.97 & 0.95 & 0.97 & 0.96 \\ 
 &  &   &  84 & 19.00 & 54.00 & 0.96 & 0.95 & 0.97 & 0.95 \\ 
 &  &   &  28 & 43.00 & 148.00 & 0.95 & 0.95 & 0.96 & 0.94 \\ 
 &  &   &  14 & 78.00 & 290.00 & 0.95 & 0.95 & 0.95 & 0.95 \\ 
 & Hotelling's $T^2_{Mp, N-q-1}$ &   & 180 & 13.00 & 29.00 & 0.96 & 0.95 & 0.97 & 0.96 \\ 
 &  &   &  84 & 19.00 & 54.00 & 0.95 & 0.95 & 0.97 & 0.96 \\ 
 &  &   &  28 & 43.00 & 148.00 & 0.95 & 0.95 & 0.96 & 0.95 \\ 
 &  &   &  14 & 78.00 & 290.00 & 0.95 & 0.95 & 0.97 & 0.95 \\ 
  \hline
\end{tabular}
\end{table}

\begin{table}[H]
\caption{Sample sizes calculation based on power (P) when the standardized proximal effect size of intervention levels satisfy $\delta_m(d)=\boldsymbol Z_{d}^{\top}\boldsymbol \delta_m$, where $\boldsymbol \delta_m=\boldsymbol\beta_m/\sigma$, for $m=1,\ldots,M_0$,$\ldots$,$\sum_{j=0}^{k}M_j$. Note that we have $k=1$, $M_0=2$ and $M_1=1$ ($M=3$), $M_0=2$ and $M_1=2$ ($M=4$), $\sigma=1$ and $\rho=0$, where $d_0=1$ and $d_1$ is the half way through ``Duration", which is the duration of study ($D$) in days, e.g., if $D=28$ then $d_1=15$. The significance level is 0.05. The desired power is 0.80. 
Constant trend for standardized proximal effect size and $100\%$ availability at each decision time point are assumed.
The initial standardized proximal effect size is 0.02.
}
\label{Table: TC7}
\centering
\begin{tabular}{rllrrrrrrr}
  \hline
   & \multicolumn{3}{c}{} &  \multicolumn{2}{c}{ Sample Size } & \multicolumn{2}{c}{ Formulated P }  & \multicolumn{2}{c}{ Monte Carlo P } \\
 & \multicolumn{3}{c}{} &  \multicolumn{6}{c}{ Average standardized proximal effect size }  \\
& Test Statistics & M & Duration & 0.20 & 0.10 & 0.20 & 0.10 & 0.20 & 0.10 \\  
  \hline
& $\chi^2_{Mp}$  & 3 & 180 &   7 &  26 & 0.84 & 0.81 & 0.84 & 0.80 \\ 
 &  &   &  84 &  14 &  55 & 0.82 & 0.81 & 0.80 & 0.81 \\ 
 &  &   &  28 &  41 & 163 & 0.80 & 0.80 & 0.79 & 0.80 \\ 
 &  &   &  14 &  82 & 325 & 0.80 & 0.80 & 0.80 & 0.80 \\ 
 & Hotelling's $T^2_{Mp, N}$ &   & 180 &  11 &  30 & 0.85 & 0.81 & 0.82 & 0.81 \\ 
 &  &   &  84 &  18 &  58 & 0.82 & 0.80 & 0.79 & 0.81 \\ 
 &  &   &  28 &  45 & 167 & 0.81 & 0.80 & 0.77 & 0.80 \\ 
 &  &   &  14 &  86 & 329 & 0.81 & 0.80 & 0.80 & 0.81 \\ 
 & Hotelling's $T^2_{Mp, N-1}$ &   & 180 &  11 &  30 & 0.82 & 0.81 & 0.80 & 0.80 \\ 
 &  &   &  84 &  18 &  59 & 0.81 & 0.81 & 0.76 & 0.77 \\ 
 &  &   &  28 &  45 & 167 & 0.80 & 0.80 & 0.80 & 0.82 \\ 
 &  &   &  14 &  86 & 329 & 0.80 & 0.80 & 0.79 & 0.80 \\ 
 & Hotelling's $T^2_{Mp, N-q-1}$ &   & 180 &  12 &  30 & 0.86 & 0.81 & 0.79 & 0.81 \\ 
 &  &   &  84 &  18 &  59 & 0.80 & 0.81 & 0.81 & 0.79 \\ 
 &  &   &  28 &  45 & 167 & 0.80 & 0.80 & 0.78 & 0.80 \\ 
 &  &   &  14 &  86 & 329 & 0.80 & 0.80 & 0.80 & 0.80 \\ 
 & $\chi^2_{Mp}$  & 4 & 180 &   7 &  28 & 0.81 & 0.81 & 0.80 & 0.79 \\ 
 &  &   &  84 &  15 &  60 & 0.81 & 0.81 & 0.82 & 0.80 \\ 
 &  &   &  28 &  45 & 178 & 0.81 & 0.80 & 0.81 & 0.78 \\ 
 &  &   &  14 &  89 & 356 & 0.80 & 0.80 & 0.82 & 0.80 \\ 
 & Hotelling's $T^2_{Mp, N}$ &   & 180 &  12 &  33 & 0.81 & 0.81 & 0.78 & 0.79 \\ 
 &  &   &  84 &  20 &  64 & 0.81 & 0.80 & 0.80 & 0.80 \\ 
 &  &   &  28 &  50 & 183 & 0.81 & 0.80 & 0.79 & 0.80 \\ 
 &  &   &  14 &  94 & 360 & 0.80 & 0.80 & 0.80 & 0.79 \\ 
 & Hotelling's $T^2_{Mp, N-1}$ &   & 180 &  13 &  33 & 0.85 & 0.81 & 0.82 & 0.81 \\ 
 &  &   &  84 &  20 &  65 & 0.80 & 0.81 & 0.77 & 0.80 \\ 
 &  &   &  28 &  50 & 183 & 0.81 & 0.80 & 0.81 & 0.79 \\ 
 &  &   &  14 &  94 & 360 & 0.80 & 0.80 & 0.76 & 0.80 \\ 
 & Hotelling's $T^2_{Mp, N-q-1}$ &   & 180 &  13 &  33 & 0.82 & 0.80 & 0.76 & 0.78 \\ 
 &  &   &  84 &  21 &  65 & 0.82 & 0.81 & 0.80 & 0.81 \\ 
 &  &   &  28 &  50 & 183 & 0.81 & 0.80 & 0.77 & 0.81 \\ 
 &  &   &  14 &  94 & 360 & 0.80 & 0.80 & 0.81 & 0.81 \\    
 \hline
\end{tabular}

\end{table}

\begin{table}[H]
\caption{Sample sizes calculation based on coverage probability when the standardized proximal effect size of intervention levels satisfy $\delta_m(d)=\boldsymbol Z_{d}^{\top}\boldsymbol \delta_m$, where $\boldsymbol \delta_m=\boldsymbol\beta_m/\sigma$, for $m=1,\ldots,M_0$,$\ldots$,$\sum_{j=0}^{k}M_j$. Note that we have $k=1$, $M_0=2$ and $M_1=1$ ($M=3$), $M_0=2$ and $M_2=1$ ($M=4$), $\sigma=1$ and $\rho=0$, where $d_0=1$ and $d_1$ is the half way through ``Duration", which is the duration of study ($D$) in days, e.g., if $D=28$ then $d_1=15$.  
The desired CP is $95\%$.
Constant trend for standardized proximal effect size and $100\%$ availability at each time point are assumed.
Precision of the initial standardized proximal effect size is 0.02.
}
\label{Table: TC8}
\centering
\begin{tabular}{rllrrrrrrr}
  \hline
  & \multicolumn{3}{c}{} & \multicolumn{2}{c}{ Sample Size } & \multicolumn{2}{c}{ Formulated CP } & \multicolumn{2}{c}{ Monte Carlo CP } \\
 & \multicolumn{3}{c}{} &  \multicolumn{6}{c}{Precision of average standardized proximal effect}  \\
& Test Statistics & M & Duration & 0.25 & 0.15 & 0.25 & 0.15 & 0.25 & 0.15 \\ 
  \hline
& $\chi^2_{Mp}$  & 3 & 180 &   5 &  19 & 0.97 & 0.96 & 0.96 & 0.95 \\ 
 &  &   &  84 &  10 &  39 & 0.96 & 0.95 & 0.95 & 0.95 \\ 
 &  &   &  28 &  30 & 117 & 0.96 & 0.95 & 0.96 & 0.96 \\ 
 &  &   &  14 &  59 & 233 & 0.95 & 0.95 & 0.95 & 0.96 \\ 
 & Hotelling's $T^2_{Mp, N}$ &   & 180 &  10 &  24 & 0.96 & 0.96 & 0.96 & 0.96 \\ 
 &  &   &  84 &  15 &  45 & 0.95 & 0.95 & 0.94 & 0.96 \\ 
 &  &   &  28 &  35 & 122 & 0.95 & 0.95 & 0.95 & 0.95 \\ 
 &  &   &  14 &  64 & 238 & 0.95 & 0.95 & 0.96 & 0.96 \\ 
 & Hotelling's $T^2_{Mp, N-1}$ &   & 180 &  10 &  24 & 0.95 & 0.95 & 0.96 & 0.97 \\ 
 &  &   &  84 &  16 &  45 & 0.96 & 0.95 & 0.97 & 0.96 \\ 
 &  &   &  28 &  35 & 122 & 0.95 & 0.95 & 0.95 & 0.96 \\ 
 &  &   &  14 &  64 & 238 & 0.95 & 0.95 & 0.95 & 0.95 \\ 
 & Hotelling's $T^2_{Mp, N-q-1}$ &   & 180 &  11 &  24 & 0.96 & 0.95 & 0.98 & 0.96 \\ 
 &  &   &  84 &  16 &  45 & 0.96 & 0.95 & 0.96 & 0.96 \\ 
 &  &   &  28 &  35 & 122 & 0.95 & 0.95 & 0.96 & 0.94 \\ 
 &  &   &  14 &  64 & 239 & 0.95 & 0.95 & 0.96 & 0.95 \\ 
 & $\chi^2_{Mp}$  & 4 & 180 &   6 &  22 & 0.97 & 0.95 & 0.96 & 0.95 \\ 
 &  &   &  84 &  12 &  48 & 0.95 & 0.95 & 0.95 & 0.96 \\ 
 &  &   &  28 &  36 & 142 & 0.95 & 0.95 & 0.95 & 0.95 \\ 
 &  &   &  14 &  71 & 283 & 0.95 & 0.95 & 0.96 & 0.95 \\ 
 & Hotelling's $T^2_{Mp, N}$ &   & 180 &  12 &  29 & 0.96 & 0.95 & 0.96 & 0.95 \\ 
 &  &   &  84 &  19 &  54 & 0.96 & 0.95 & 0.97 & 0.95 \\ 
 &  &   &  28 &  42 & 148 & 0.95 & 0.95 & 0.96 & 0.94 \\ 
 &  &   &  14 &  78 & 290 & 0.95 & 0.95 & 0.95 & 0.96 \\ 
 & Hotelling's $T^2_{Mp, N-1}$ &   & 180 &  13 &  29 & 0.97 & 0.95 & 0.97 & 0.95 \\ 
 &  &   &  84 &  19 &  54 & 0.96 & 0.95 & 0.96 & 0.96 \\ 
 &  &   &  28 &  43 & 148 & 0.95 & 0.95 & 0.95 & 0.96 \\ 
 &  &   &  14 &  78 & 290 & 0.95 & 0.95 & 0.96 & 0.96 \\ 
 & Hotelling's $T^2_{Mp, N-q-1}$ &   & 180 &  13 &  29 & 0.96 & 0.95 & 0.97 & 0.96 \\ 
 &  &   &  84 &  19 &  54 & 0.95 & 0.95 & 0.98 & 0.95 \\ 
 &  &   &  28 &  43 & 148 & 0.95 & 0.95 & 0.95 & 0.95 \\ 
 &  &   &  14 &  78 & 290 & 0.95 & 0.95 & 0.95 & 0.94 \\       
   \hline
\end{tabular}
\end{table}

\end{document}